\documentclass[apj,twocolappendix]{emulateapj}

\usepackage{mathtools}
\usepackage{graphicx}
\usepackage[bottom]{footmisc}
\usepackage{mathptmx}
\usepackage{color}
\usepackage[breaklinks=true, colorlinks=true, linkcolor=red, urlcolor=blue, citecolor=black]{hyperref}
\definecolor{myblue}{RGB}{0, 0, 150}

\def\Mpc{{\rm Mpc}}
\def\kpc{{\rm kpc}}

\def\kms{{\rm km}\,{\rm s}^{-1}}

\def\Msun{{\rm M}_\odot}
\def\Mstel{M_\ast}

\def\logM{\log\Mstel/\Msun}

\def\resp{respectively}
\def\bfr{\bf\color{red}}
\def\bfb{\color{black}}

\def\ssfr{{\rm sSFR}}
\def\sfr{{\rm SFR}}

\def\dens{\Sigma_{e}}
\def\zphot{z_{\rm phot}}
\def\zspec{z_{\rm spec}}

\def\beq{\begin{equation}}
\def\eeq{\end{equation}}
\def\bitem{\begin{itemize}}
\def\eitem{\end{itemize}}
\def\benum{\begin{enumerate}}
\def\eenum{\end{enumerate}}

\mathchardef\mhyphen="2D

\def\cite{{\bfr CITE}}

\def\compl{9.4}
\def\ntot{1491}
\def\nred{233}
\def\nblue{1258}

\shortauthors{Abramson \& Morishita}
\shorttitle{The Weak Evidence for Compaction}

\begin{document}

\title{
Must {\bfb Starforming} Galaxies Rapidly Get Denser Before They Quench?
}

\slugcomment{Accepted to ApJ, 8 March 2018}

\author{
L.E.~Abramson\altaffilmark{1} and
T.~Morishita\altaffilmark{2}
}


\begin{abstract}

Using the deepest data yet obtained, we find no evidence preferring compaction-triggered quenching---where rapid increases in galaxy density truncate star formation---over a null hypothesis in which galaxies {\bfb age} at constant surface density ($\dens\equiv\Mstel/2\pi r_{e}^{2}$). Results from two fully empirical analyses and one quenching-free model calculation support {\bfb this claim at all $z\leq3$:} (1) Qualitatively, galaxies' mean $U-V$ colors at $6.5\lesssim\log\dens/\Msun\,\kpc^{-2}\lesssim10$ have reddened at rates/times correlated with $\dens$, implying that there is no density threshold at which galaxies turn red but that $\dens$ sets the pace of maturation; (2) Quantitatively, the abundance of $\logM\geq9.4$ red galaxies never exceeds that of the total population a quenching time earlier at any $\dens$, implying that galaxies need not transit from low to high densities before quenching; (3) Applying $d\log r_{e}/dt =1/2\,d\log\Mstel/dt$ to a suite of lognormal star formation histories reproduces the {\bfb evolution of the} size--mass relation at $\logM\geq10$. All results are consistent with evolutionary rates being set {\it ab initio} by global densities, with denser objects evolving faster than less-dense ones towards a terminal quiescence induced by gas depletion or other $\sim$Hubble-timescale phenomena. Unless stellar ages {\bfb demand} otherwise, observed $\dens$ thresholds need not bear any physical relation to quenching beyond this intrinsic density--formation epoch correlation, adding to Lilly \& Carollo's arguments to that effect.

\end{abstract}


\altaffiltext{1}{
UCLA, 430 Portola Plaza, Los Angeles, CA 90095-1547, USA; \href{mailto:labramson@astro.ucla.edu}{labramson@astro.ucla.edu}
}
\altaffiltext{2}{
Space Telescope Science Institute, 3700 San Martin Drive, Baltimore, MD 21218, USA
}


\section{Context}
\label{sec:intro}

Why some galaxies form stars while others do not is a central puzzle in astronomy. Quiescence correlates with mass, environment, kinematics, and structure, but whether/how these factors cause the cessation of star formation ({\bfb ``quenching''}) is unknown. The same holds for stellar mass density.

For at least half a century, non-starforming galaxies have been known to be denser than contemporaneous, equal-mass starforming ones \citep{Holmberg64}. Recent studies amplify/extend this finding \citep[][]{Fang13, Barro13, Barro15, Whitaker17}, but do not clarify its meaning. As \citet[][``LC16'']{Lilly16} discuss, there are causal and corollary interpretations.

The causal scenario is that starforming galaxies experience dramatic (gas) density increases due to mergers/instabilities---``compaction.'' This leads to strong, concentrated starbursts whose outflows/subsequent black hole activity stifle further star formation, leaving {\bfb small}, passive stellar cores \citep[][]{Barro15, Zolotov15}. {\bfb Violent} density {\bfb increases beyond some stability threshold} thus {\it trigger} {\bfb quenching}.

The corollary scenario is that {\it ab initio} denser galaxies form stars faster than less-dense ones \citep[][]{Schmidt59, Kennicutt98}. This accelerated evolution leads them to quiescence---galaxies' natural end-state---first, via gas exhaustion or other Hubble-timescale processes (which are by definition rapid at high-$z$). With no {\bfb pre-quenching increases} at all, passive systems {\bfb would be} denser than starforming ones{\bfb, and denser systems would be more passive---as the size--mass plane shows (\citealt{Holmberg64, Dressler80, vanDerWel14, Abramson16}; LC16; \citealt{Morishita17}).}

{\bfb The question for quenching physics is whether the data {\it require} something beyond an intrinsic density--formation time covariance \citep[akin to assembly bias;][]{Wechsler06,Feldmann16}.} Based on their analysis of a model in which galaxies grow via scaling laws in mass, $\ssfr$($\equiv\sfr/\Mstel$), and time until {\bfb quenched} by a mass- {\bfb or environment-}sensitive mechanism, LC16 suggest ``no''---{\bfb data do not favor} causality (compaction) over correlation (denser things die first). Here, {\bfb we present complementary, concurring arguments based on a more extreme null hypothesis: From both fully empirical and independent model analyses, we find no compelling reason to dismiss a scenario wherein pre-quenching galaxy surface densities {\it never evolve}. Hence, either the} size--mass relation is a poor test of physical models, or $\dens\propto\Mstel/r_{e}^{2}$ {\bfb is a plausibly} conserved quantity for blue galaxies.

{\bfb Below, Section \ref{sec:data} describes the data, \ref{sec:dataResults} analyzes them in isolation, and \ref{sec:toyModel} provides separate supportive modeling. Section \ref{sec:discussion} discusses counterarguments, and Appendices \ref{sec:AA}--\ref{sec:AC} present cross-checks. Please see \ref{sec:clarification} before interpreting any results.}

We assume $(H_{0},\Omega_{m},\Omega_{\Lambda}) = (73\,\kms\,\Mpc^{-1},0.27,0.73)$.


\begin{figure*}[t!]
\centering
\includegraphics[width = 0.425\linewidth, trim = 2.5cm 0.5cm 0cm 0cm]{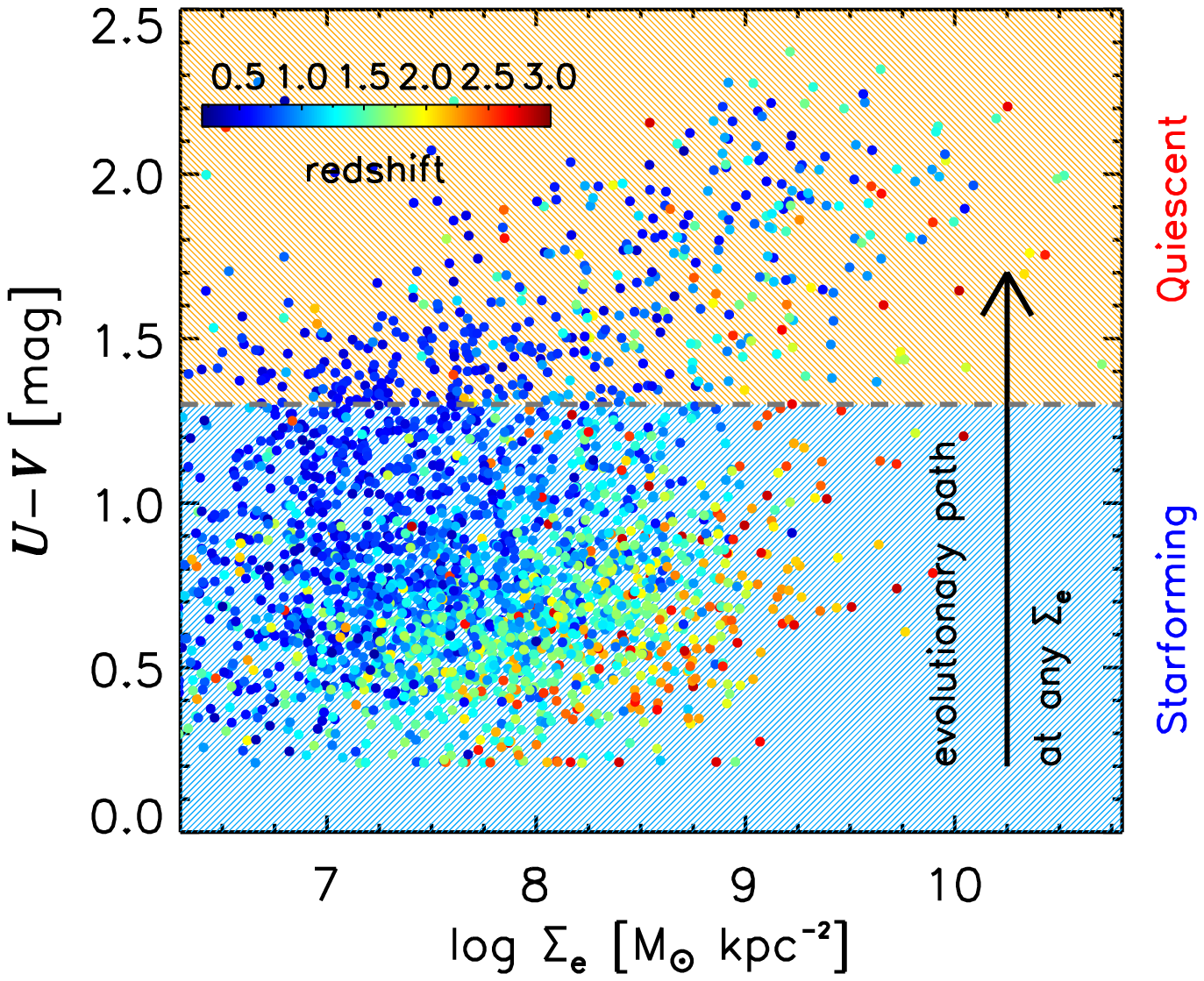}
\includegraphics[width = 0.425\linewidth, trim = 0.5cm 0.5cm 2cm 0cm]{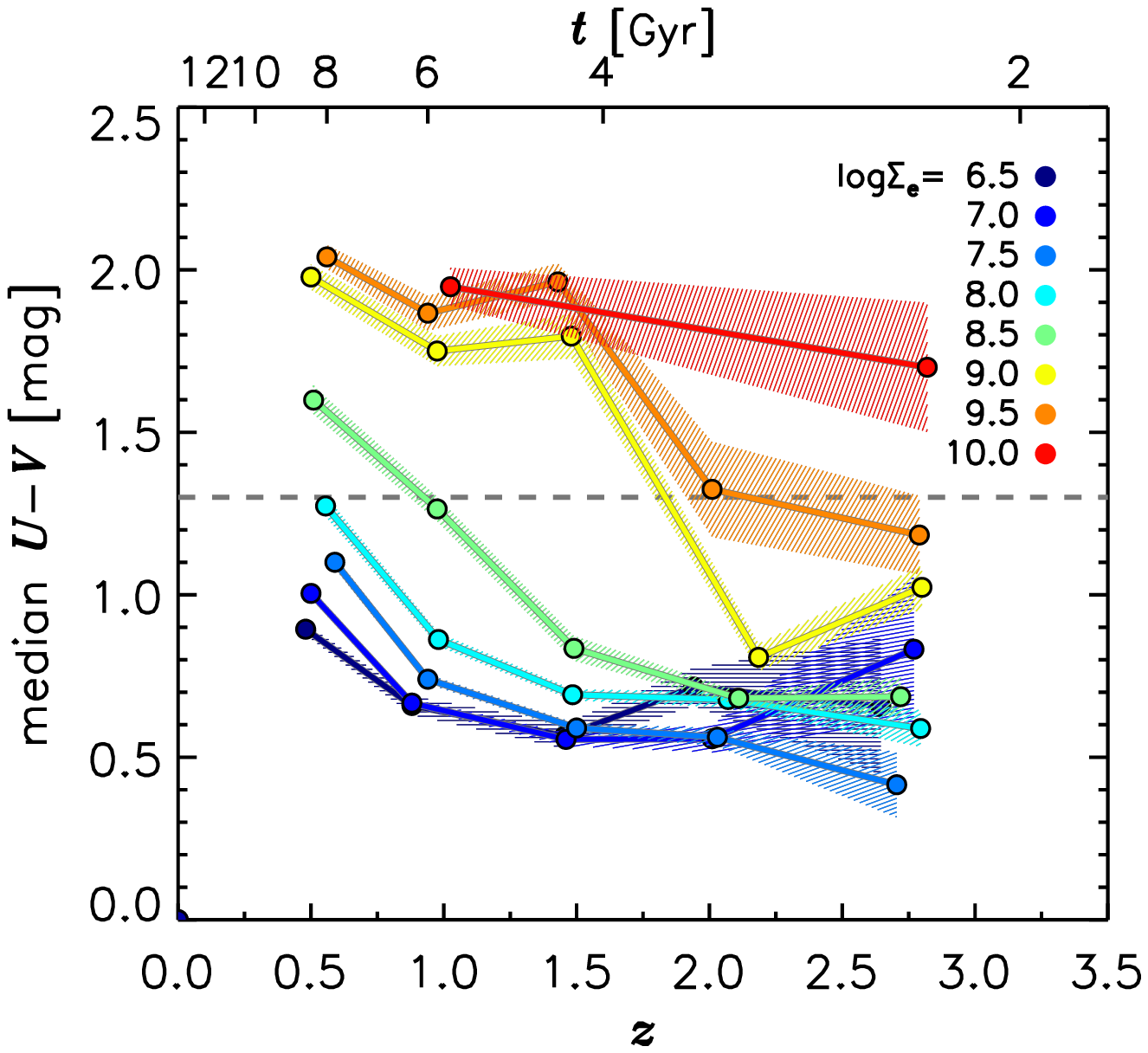}
\caption{Colors are consistent with blue galaxies evolving into red ones at fixed mass surface density {\it Left}: Higher-$\dens$ objects seem to quench first, but could likewise age fastest, their natural evolution to more massive, passive systems speeded by globally higher gas densities at earlier times (\citealt{Holmberg64}, Figure 6; \citealt{Franx08}, Figure 6). {\it Right}: Galaxies at {\it all} $\log\dens\geq6.5$---$\sim$100$\times$ below \citet{Zolotov15}'s quenching range---have reddened over time with the transition slowed/delayed at lower densities. No $\dens$ threshold is seen; galaxies of different densities may simply age according to different clocks \citep{Gladders13b,Abramson16}. Note: $U-V$ color is plotted as a qualitative metric. Quantitative analyses use the full {\it UVJ} quiescent/starforming classification criteria (Section \ref{sec:data}). Bands at {\it left} show 1\,$\sigma$ color scatter divided by $\sqrt{N_{\rm gals}(\dens; z)}$.}
\label{fig:densEvo}
\end{figure*}

\section{Data}
\label{sec:data}

Our empirical analysis uses public {\it Hubble Space Telescope} (HST) data from the {\it eXtreme Deep Field} \citep[XDF;][]{Illingworth13},\footnote{\url{https://archive.stsci.edu/prepds/xdf}} {\it Hubble Legacy Fields} \citep[HLF;][]{Illingworth16},\footnote{\url{https://archive.stsci.edu/prepds/hlf}} {\it Hubble Frontier Fields} \citep[HFF;][]{Lotz17},\footnote{\url{https://archive.stsci.edu/prepds/frontier}} and the {\it Grism Lens-Amplified Survey from Space} \citep[GLASS;][]{Treu15}.\footnote{\url{https://archive.stsci.edu/prepds/glass}} This imaging and spectroscopy represents the deepest data yet obtained, probing hitherto inaccessible stellar mass and redshift regimes ($\log\Mstel\gtrsim9$ at $z\lesssim3$).

We take the 7-band (F435/606/814/105/125/140/160W) imaging covering the six HFF cluster and parallel pointings, and the 9-band imaging (HFF + F775W + F850LP) from the XDF and two HLF fields with comparable F160W data (used for structural fitting). To {\bfb ensure} environmental and structural effects {\bfb are not confused}, we exclude HFF/GLASS cluster members {\bfb \citep[see][``M17'']{Morishita17}.}

Source detection, and photometric redshift ($\zphot$), rest-frame color, and stellar mass estimation follow M17: {\tt EAZY} \citep{Brammer08} determines $\zphot$ using M17's modified prior to identify cluster members; {\tt FAST} \citep{Kriek09} yields stellar masses based on $\zphot$ or GLASS $\zspec$ (if available) assuming a \citet{Chabrier03} IMF, \citet{Calzetti00} dust law, and exponential {\bfb star formation} history (SFH).

{\bfb {\tt GALFIT} \citep{PengGALFIT} provides} galaxy sizes---circularized F160W effective radii; $r_{e}\equiv a_{e}\sqrt{q}$, {\bfb where $q$ is the axis ratio and $a_{e}$ is the major-axis} half-light radius---assuming a \citet{Sersic63} profile. We study systems with $r_{e}\ge{\rm HWHM_{F160W}}=0\farcs09$, which are well-resolved \citep{Morishita14}. Bright stars serve as point spread functions.

We identify blue/starforming and red/non-starforming galaxies using the {\it UVJ} color-color criteria of \citet[][see their Equation 4]{Williams09}, cutting the sample at $m_{\rm F160W}\leq26$ AB where structural fits are reliable (M17). This is 1.5~mag deeper than \citet[][``vdW14'']{vanDerWel14}'s CANDELS data. Scaling their mass completeness limit---$\log\Mstel=10.0$ for $z\sim2.5$ red galaxies---our data reach $\log\Mstel\sim9.4$ at $0.2\leq z\leq3$. Results are robust to this limit {\bfb (Appendix \ref{sec:AA})}. 

The sample contains \ntot\ (\nred\ red + \nblue\ blue) galaxies.

\subsection{A Note on $\Sigma_{e}$}

Though some authors favor $\Sigma_{1}$ or $\rho_{1}$---the surface or 3D stellar mass density at $r\leq1$ kpc \citep[][]{Fang13,Barro15,Whitaker17}---we adopt $\dens\equiv\Mstel/2\pi r_{e}^{2}$ because: (1) it is less sensitive to fitting errors; (2) we lack spatially resolved colors \citep[using a global $\Mstel/L$ is likely the dominant systematic;][]{Szomoru13, Morishita15}; (3) it avoids additional assumptions needed to infer $\Sigma_{1}$ or $\rho_{1}$ from 2D profiles; (4) we wish to test a null scenario where galaxies evolve at constant surface density but obviously not constant mass. As star formation is often uniformly distributed \citep{Nelson16}, this precludes fixed-aperture density definitions, where $d\Sigma/dt$ effectively becomes $d\Mstel/dt$. As a penalty, we must account for differential size/mass effects (Sections \ref{sec:toyModel}, \ref{sec:discussion}), and verify that $\Sigma_{1}$ is consistent for $\dens$-matched progenitors/descendants (Section \ref{sec:advancedResults}).


\section{Empirical Results: Colors, Abundances}
\label{sec:dataResults}

{\bfb
\subsection{Clarification of Intent}
\label{sec:clarification}

Before discussing our results, we wish to emphasize their meaning. We seek to assess whether galaxy sizes, masses, and colors imply rapid, quenching-related $\dens$ increases. We claim they do not, and argue by showing that a null hypothesis of constant-$\dens$ evolution also fits the facts. {\it We do not claim that galaxies, in reality, should grow at constant $\dens$}. 

This section's empirical analyses satisfy our aim: no data suggest anything beyond constant-$\dens$ aging. One can view this as evidence for the physicality of that scenario, or a measure of the data's constraining power. We are agnostic.

Section \ref{sec:toyModel}'s modeling is different. It can only {\it affirm} that {\it all galaxies might actually grow} at constant $\dens$. To uniquely test that much stronger statement---which is not our intent---the external SFHs and boundary conditions must be correct. Of course, in general, they will not be. As such, the test is unidirectional: consistency with data suggests constant-$\dens$ growth is not wrong, but inconsistency does not suggest it is---the issues could lie with the SFHs or boundary conditions. 

As it stands, our most-basic model accounts for $>$90\% of $\log\Mstel>10$ galaxies at $z<3$ (Appendix \ref{sec:AB}), and Appendix \ref{sec:AC} illustrates how it could be modified to account for the rest. We take this as sufficient support for our empirical arguments, but encourage further tests by any who disagree.} 

\subsection{\bfb Colors}

Figure \ref{fig:densEvo}, {\it left}, shows the data: galaxy colors as a function of stellar mass surface density---$\dens\equiv\Mstel/2\pi r_{e}^{2}$---and redshift. Starforming galaxies occupy the entire $\dens$ range at high-$z$, but systematically vacate the high-density end as $z\rightarrow0$. Conversely, high-$z$ red galaxies lie almost exclusively at the highest $\dens$, but appear later at lower {\bfb density}. These statements are not inconsistent with galaxies evolving in an L-shaped track \citep{Barro15, Zolotov15}, increasing in density (moving right $\rightarrow$ ``blue nuggets'') before reaching a redshift-dependent, monotonically decreasing threshold and quenching (moving up $\rightarrow$ ``red nuggets''). 

{\bfb {\it Yet}, these facts} are {\bfb also} consistent with blue galaxies providing a source population spanning all $\dens$ that is systematically emptied as they age into red systems at relatively fixed surface density; {\bfb i.e.,} $d\dens/dt\ll d\ssfr/dt$ \citep[][Figure 6]{Holmberg64}. If, reasonably, such evolution was fastest for the densest systems---{\bfb i.e.,} the earliest, assuming galaxies encode global densities at some formative epoch (Section \ref{sec:toyModel})---then this ``peeling'' scenario would appear as {\bfb $\dens$-threshold-dependent} quenching but entail no mechanism beyond dense gas supporting higher SFRs than rarefied gas (\citealt{Dressler80, Kennicutt98, Poggianti13b, Abramson16}, LC16, \citealt{Kelson16}; M17). {\it Irrespective of its ultimate accuracy, we see this as a null scenario requiring falsification}.

Figure \ref{fig:densEvo}, {\it right}, summarizes: galaxies at all $\dens$ have reddened with time. This process is merely delayed or retarded for lower- compared to higher-density objects. A mass dependence is embedded in these results in that lower-$z$ galaxies are more massive than higher-$z$ systems of equal $\dens$, but this is the point---galaxies might grow in mass at constant $\dens$. 

Figure \ref{fig:densEvo} is {\bfb thus qualitatively} consistent with evolutionary rates being set {\bfb by an initial density}, then remaining fixed: Denser galaxies born earlier might experience the same physical processes as less-dense galaxies born later, just sooner/at accelerated rates with no additional $\dens$-dependent process required (\citealt{Gladders13b,Kelson14,Papovich15,Abramson16}; LC16; Section \ref{sec:toyModel}). In such a universe, {\bfb a sharp, evolving} $\dens$ ``threshold'' {\bfb could still separate blue and red galaxies} (LC16; \citealt{Whitaker17}). Yet, rather than a {\bfb quenching} mechanism, the signal would reflect the fact that all higher-$\dens$ galaxies have ``aged-out'' of the starforming population and are thus not present in lower-$z$ blue samples to support the measurement.

Since no $\dens$ is preferred for the reddening process---which can take many Gyr---a {\bfb plain} interpretation {\bfb of Figure \ref{fig:densEvo}} is that there is no $\dens$ {\bfb threshold-triggered quenching}. {\bfb Of course}, the data {\bfb also admit} an evolving quenching threshold.

\begin{figure*}[t!]
\centering
\includegraphics[width = 0.475\linewidth, trim = 0.5cm 0cm -0.5cm 0cm]{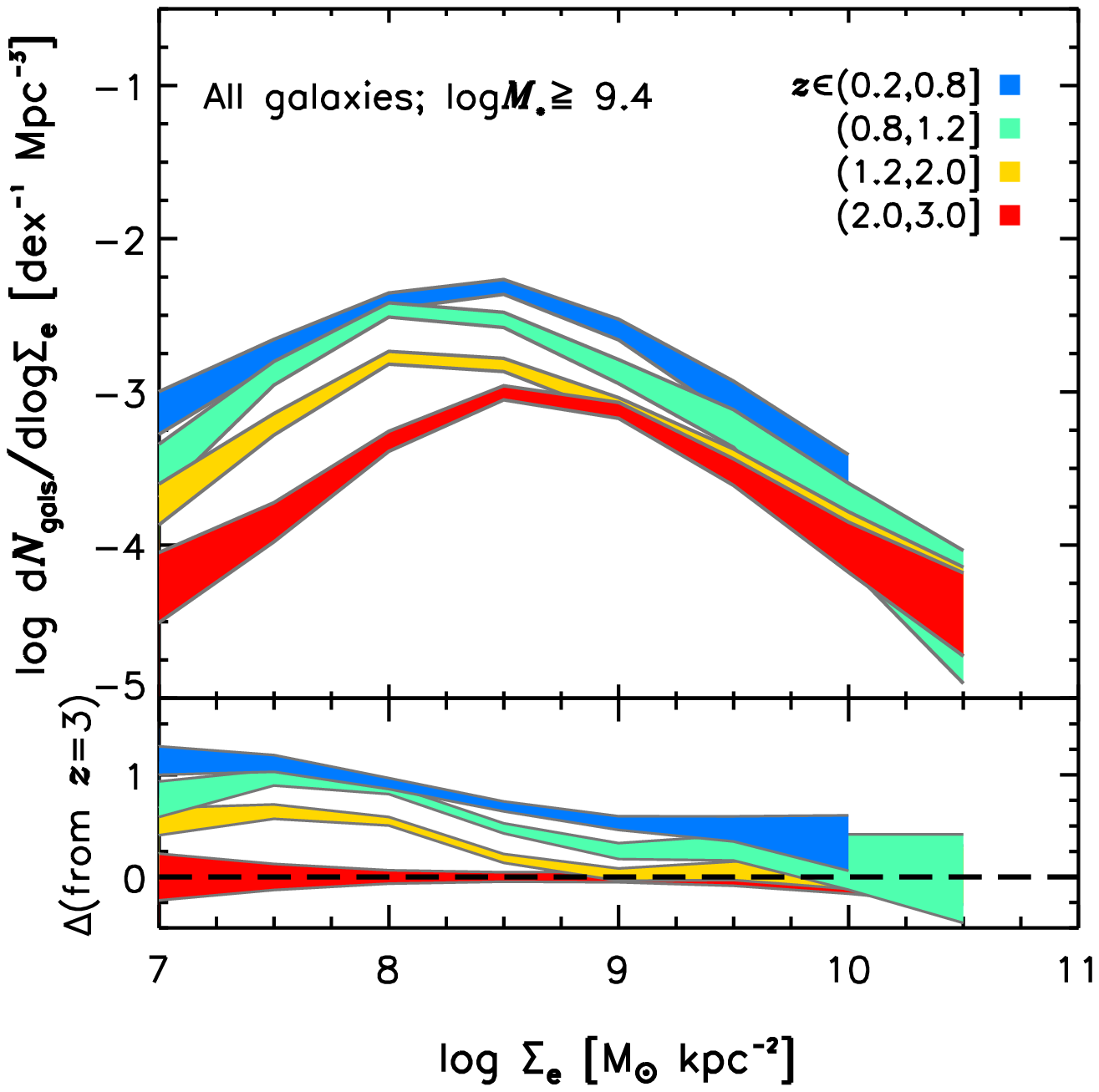}
\includegraphics[width = 0.475\linewidth, trim = -0.5cm 0cm 0.5cm 0cm]{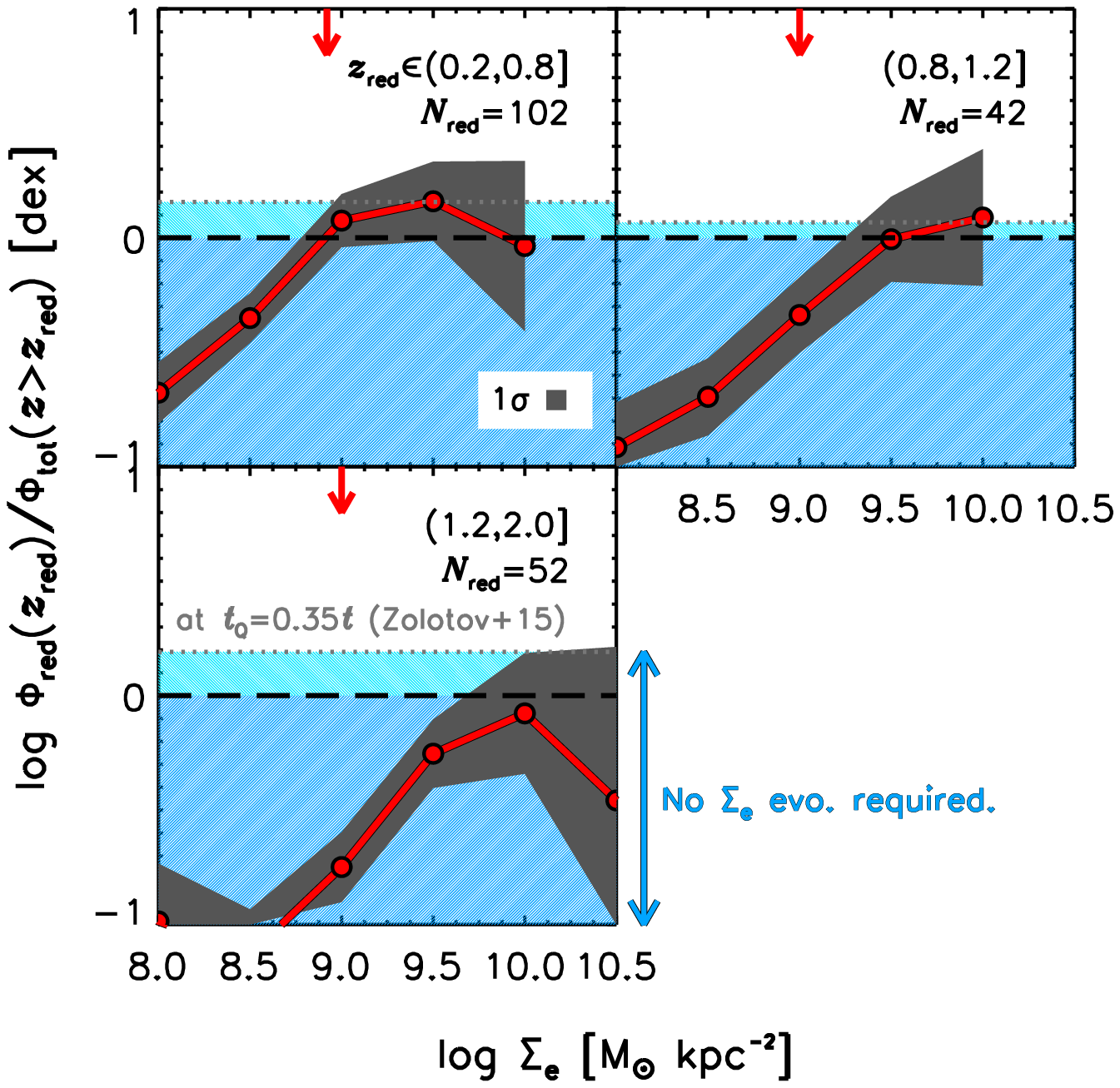}
\caption{{\it Left}: At $z\leq3$, galaxy counts grow faster at lower $\dens$  (quantified at {\it bottom}), limiting how many objects can move to higher $\dens$ (left-to-right) over time. If such systems quenched by crossing a $\dens$ limit, {\it UVJ} passive galaxy counts at lower $z_{\rm red}$ could exceed those of all equal-$\dens$ higher-$z$ objects. {\it Right}: The data do not show this signal. Even above published thresholds \citep[red arrows;][]{Whitaker17}, no significant overabundances are seen, especially when considering that $z$ intervals exceed a \citet{Zolotov15} quenching time ($\tau_{\rm Q}=0.35\,t_{z}$), allowing galaxies to enter the sample and quench between measurements (dotted grey lines denote $\log[(t_{z_{\rm red}} - t_{z})/\tau_{\rm Q}]$; Section \ref{sec:advancedResults}). {\bfb Appendix \ref{sec:AA}, Figure \ref{fig:CANDELS} verifies these results using vdW14's larger, shallower sample.}} 
\label{fig:histograms}
\end{figure*}

\subsection{\bfb Abundances}
\label{sec:advancedResults}

{\bfb The above} results favor neither the corollary density-accelerated scenario, nor the causal compaction/density-quenched model. Here, we perform a test that increases the burden of proof on the latter.
{\bfb Appendix \ref{sec:AA} verifies all results in this section using the wider-but-shallower vdW14 sample.}

Figure \ref{fig:histograms}, {\it left}, shows the evolution of the $\dens$ function---the absolute abundance of galaxies at a given surface density---over $0.2\leq z\leq3.0$ for all systems with $\log\Mstel\ge\compl$, the $z\sim3$ completeness limit. Uncertainties are the quadrature sum of Poisson error and 100 random perturbations by $\dens$ error-bars. 

\begin{figure}[t!]
\centering
\includegraphics[width = \linewidth, trim = -0.5cm 0cm 0cm 0cm]{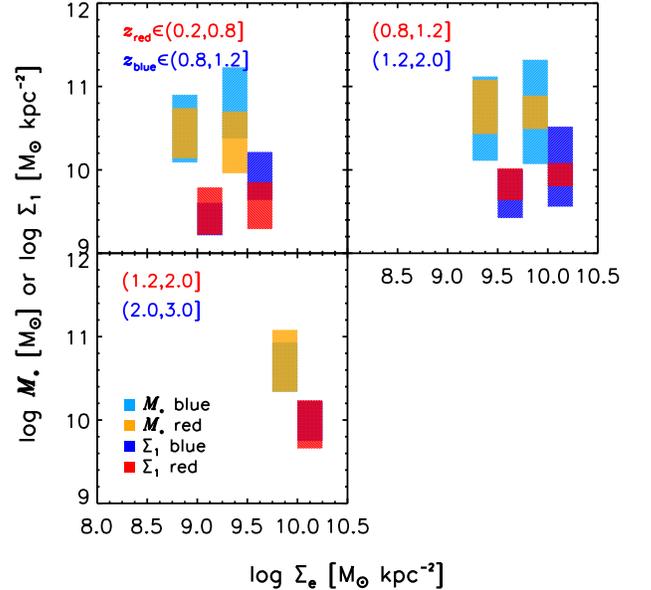}
\caption{The $16^{\rm th}$--$84^{\rm th}$ percentiles in $\Mstel$ and $\Sigma_{1}$ for blue galaxies at $z>z_{\rm red}$ and red ones at $z_{\rm red}$ overlap significantly at $\dens$s where quenching is prominent (Figure \ref{fig:histograms}).  Hence, beyond abundances, the detailed properties of many higher-$z$ blue galaxies are consistent with lower-$z$, equal-$\dens$ red ones. {\bfb Appendix \ref{sec:AA}, Figure \ref{fig:sig1CheckCANDELS} verifies this using vdW14's larger, shallower sample.}}
\label{fig:sig1Check}
\end{figure}

In agreement with \citet{Poggianti13, Poggianti13b}, at $\sim$2\,$\sigma$, nearly all $\log\dens\gtrsim10$ galaxies appear by $z\sim3$. Below that, abundances grow {\bfb faster} with decreasing $\dens$. This is attributable to the evolution of the blue galaxy mass function \citep[e.g.,][]{Ilbert10} and size--mass relation; {\bfb i.e., more and larger systems crossing the {\it mass}---not $\dens$---completeness limit as $z\rightarrow0$}  (e.g., vdW14; LC16). {\bfb Indeed,} the slope and dispersion of the size--mass relation allow equal-$\dens$ (-$\Sigma_{1}$) galaxies to span $\sim$30$\times$ (10$\times$) in $\Mstel$ \citep[][M17; {\bfb Figures \ref{fig:sig1Check}, \ref{fig:sig1CheckCANDELS}}]{Barro15}. Section \ref{sec:toyModel} illustrates the far-reaching consequences of this fact.

With colors, we can interpret these abundances to constrain the {\it need} for {\bfb compaction-triggered} quenching:
\bitem
	\item {\bf If galaxies age at constant $\dens$}, then red {\bfb systems} are the memory of once-blue ones in the same density bin. Hence, {\it the $\dens$ function of red galaxies should never exceed that of all galaxies a quenching time ago}. (There cannot be more descendants than progenitors.)
	\item {\bf If galaxies evolve strongly in $\dens$}, then the above need not be true: previously low-$\dens$ blue galaxies can compactify to end-up as high-$\dens$ red ones. Hence, {\it the $\dens$ function of red galaxies can exceed that of all earlier galaxies}: the red population may draw progenitors from their $\dens$ bin and the large reservoir of lower-$\dens$ systems.
\eitem
{\bfb Modulo mergers (Section \ref{sec:discussion}), s}izable overabundances of lower-$z$ red galaxies compared to all higher-$z$, equal-$\dens$ galaxies would therefore indicate compaction-triggered quenching.

{\bfb We search for these following} \citet{Wild16}. Figure \ref{fig:histograms}, {\it right}, compares the red galaxy $\dens$ functions at $\langle z_{\rm red}\rangle\in\{0.5,1.0,1.6\}$ to the total $\dens$ functions at $\langle z\rangle\in\{1.0,1.6,2.5\}$. Zero is the expectation if all galaxies in a $\dens$ bin at $z_{i}$ quenched by $z_{{\rm red},i}$ with no systems {\bfb added} between intervals. {\bfb Hence, in an infinitely complete sample (explained momentarily), u}nder strict constant-$\dens$ evolution, the red lines cannot lie significantly above the black dashes. 

At all densities and times except potentially one $z\sim0.5$ bin, this is precisely what is seen: Red galaxy counts never exceed the those of all older equal-$\dens$ galaxies{\bfb, consistent with constant $\dens$ expectations}. 

Varying $z$-intervals and bin sizes, we can create $\lesssim$2.5\,$\sigma$ tension at $z<1$. {\bfb Yet, even ignoring that compaction is thought to be subdominant at these epochs for spectrophotometric reasons (\citealt{Yano16}; but cf.\ \citealt{Wild16}), this elevation {\it with respect to zero} need not imply variable-$\dens$ growth. This is because the sample is {\it not} infinitely complete, and there is a source term of blue galaxies at many $\dens$ (Figure \ref{fig:densEvo}). Hence,} if $z$-intervals are longer than a quenching time---$\tau_{\rm Q} = 0.35\,t(z)\sim0.9$--2.0 Gyr \citep{Zolotov15}---{\bfb excesses could indicate those galaxies crossing the survey's {\it mass} limit,} entering the sample, and quenching between measurements.\footnote{Shorter/longer $\tau_{\rm Q}$ will increase/decrease this leeway. \citet{Abramson16} find $\tau_{\rm Q}\sim0.2\,t$.}  Indeed, {\bfb the total abundance of passive galaxies} only exceeds {\bfb that of all $z\sim2.5$ galaxies} at $z\lesssim0.9$, {\bfb providing} an ample $\gtrsim$3.5\,Gyr for such an influx {\bfb given typical mass doubling times at those epochs \citep[$\langle\ssfr\rangle^{-1}\sim0.5$--1\,Gyr;][]{Noeske07,Daddi07}.} Accounting for the extra $t(z_{\rm red}) - t(z)>\tau_{\rm Q}$ time removes all tension (grey dotted lines).

{\bfb An implication of the above is that the evolution in Figure \ref{fig:histograms}, {\it left}, purely reflects mass incompleteness. As such, {\it JWST} should provide strong tests: the undetected progenitors we posit will appear in its surveys. Under constant-$\dens$ growth, the $\dens$ function at all epochs should then resemble $z\sim0$ data (at least in regimes where dry merging is rare). If not, our strictest null scenario would be ruled out. Current data may already permit testing in this vein  \citep[e.g.,][]{Bouwens17}.}

{\bfb Of course,} {\it mild} $\dens$ {\bfb evolution} could also {\bfb occur}. Though it violates the strictest null scenario, a process where $d\dens/dt\ll d\ssfr/dt$ seems distinct from compaction-triggered quenching \citep[cf.][Figures 2, 3]{Zolotov15}. Since $z\sim0.5$ red galaxy abundances are within a factor of 2 of all $z\sim2.5$ galaxies' ($\log\dens\geq8$), $\dens$-bin crossing rates of $1/\Delta t\sim0.2$ Gyr$^{-1}$ could explain the excess. This is slow compared to $1/\tau_{\rm Q}$ \citep{Poggianti13b}, except at $z\lesssim0.5$, where, {\bfb again}, compaction is {\bfb thought to be a weak channel}. Further, progenitors of (e.g.) Milky Way-mass objects may grow by $\gtrsim$10$\times$ in $\Mstel$ over this interval \citep[][]{Leitner12,Abramson15}, emphasizing mass- over $\dens$-driven quenching (LC16).

Although abundances match, more detailed properties of equal-$\dens$ galaxies at various times---e.g., $\Sigma_{1}$, $\Mstel$---might not, ruling out (quasi-)fixed-$\dens$ evolution. Yet, when $\Sigma_{1}$ is inferred from 1D projections of the {\tt GALFIT} models \citep{Bezanson09}, Figure \ref{fig:sig1Check} (and \ref{fig:sig1CheckCANDELS}) shows significant overlap between these quantities for earlier blue and later red galaxies at fixed $\dens$. Hence, no strong tension emerges at the galaxy level.

\section{Numerical Results: the Size--Mass Relation}
\label{sec:toyModel}

{\bfb The empirical tests above {\bfb (and in Appendix \ref{sec:AA})} self-consistently show that the data are explainable by constant-$\dens$ evolution and do not require compaction. Yet, they do not show that, given realistic SFHs, constant-$\dens$ growth generates the data. Here, we provide support for this using a basic forward-model of the $z\lesssim3$ size--mass relation based on \citet[][``G13'']{Gladders13b}'s well-tested SFHs. As we seek only to assess whether constant-$\dens$ growth is acceptable---not prove it is uniquely good or rigorously constrain the ``true'' $r_{e}(\Mstel;\,t)$ scaling (Section \ref{sec:clarification})---we leave probing SFH-dependence to future work. Nevertheless, a basic G13+constant-$\dens$ model explains the data well (Appendix \ref{sec:AB}), and Appendix \ref{sec:AC} shows how it might readily be improved. Section \ref{sec:dataResults}'s results stand independent of any modeling.

\subsection{The Model}
\label{sec:modelIngredients}

Two features of the size--mass relation appear challenging. First, the locus for starforming galaxies follows $\log r_{e}\propto0.1$--0.3\,$\log\Mstel$ (\citealt{Shen03, vanDerWel14, vanDokkum15}; M17), shallower than the 0.5 dex/dex slope constant-$\dens$ growth implies. Second, compaction posits that small, high-mass (passive) objects come from once-{\it larger}, lower-mass systems \citep{Barro15}. The constant-$\dens$ scenario demands smaller progenitors, which, depending on the epoch, may correspond to unphysical sizes.}

{\bfb To check the above, we require} a set of SFHs and initial sizes: the former provide $d\Mstel/dt$ with which to update the latter by $d\log r_{e}=1/2\,d\log\Mstel$ to create constant-$\dens$ size trajectories. We adopt G13's lognormal SFHs. These reproduce many properties of the galaxy population{\bfb, including the total stellar mass function at $z\lesssim8$, those of starforming and passive galaxies at $z\lesssim2.5$, and the $\sfr$--$\Mstel$ relation at $z\lesssim7$ (G13; \citealt{Abramson15,Abramson16}). They also well-approximate simulated {\bfb and observationally inferred} SFHs \citep{Diemer17a, Dressler16, Abramson17b}.} As G13 employs no global scaling laws, {\bfb this analysis represents a meaningful, independent complement to LC16's}.

We generate initial galaxy sizes, $r_{e,0}$, by picking a stellar mass, $M_{*,0}$, at which galaxies ``learn'' about the density of the Universe, and assigning each galaxy:
\begin{equation}
	r_{e,0} = \eta\left[\frac{M_{*,0}}{\rho_{200c}(z_{0})}\right]^{1/3} {\rm kpc},
\label{eq:r0}
\end{equation}
where $\rho_{200c}(z_{0})$ is 200$\times$ the critical density at the redshift when a galaxy reaches $M_{*,0}$ [$\rho_{c}=(1+z_{0})^3\,3\,\Omega_{m}\,H_{0}^{2}/8\pi G$]. ${\rm Log}\,\eta=1.25$ is a constant encoding the mean, e.g., stellar-to-halo mass ratio, offset between halo $r_{200c}$ and the quantity to which stellar structures are actually sensitive \citep{Diemer17b, Baxter17}, and angular momentum and dissipation effects. To account for some diversity in the above, we {\it randomly} perturb $r_{e,0}$ using a gaussian with $\sigma=0.2$ dex---roughly the scatter in angular momenta at fixed halo mass \citep[][]{MoMaoWhite98,Burkert16}. This was derived independently from, but concurs with LC16.

\begin{figure*}[t!]
\centering
\includegraphics[width = 0.95\linewidth, trim = 1.5cm -0.25cm 0cm 0.5cm]{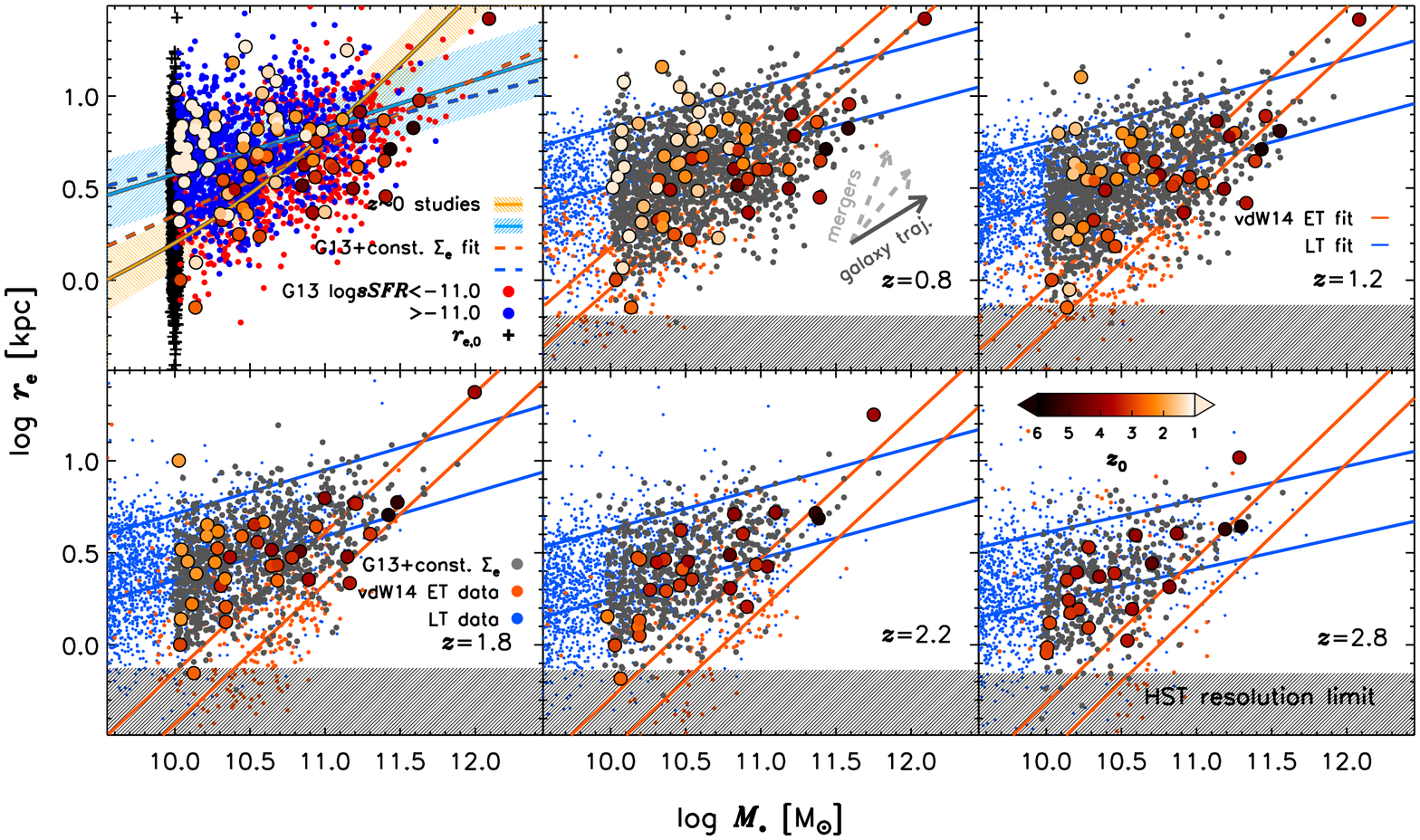}\\
\includegraphics[width = 0.95\linewidth, trim = 1.5cm -0.5cm 0cm 0.75cm]{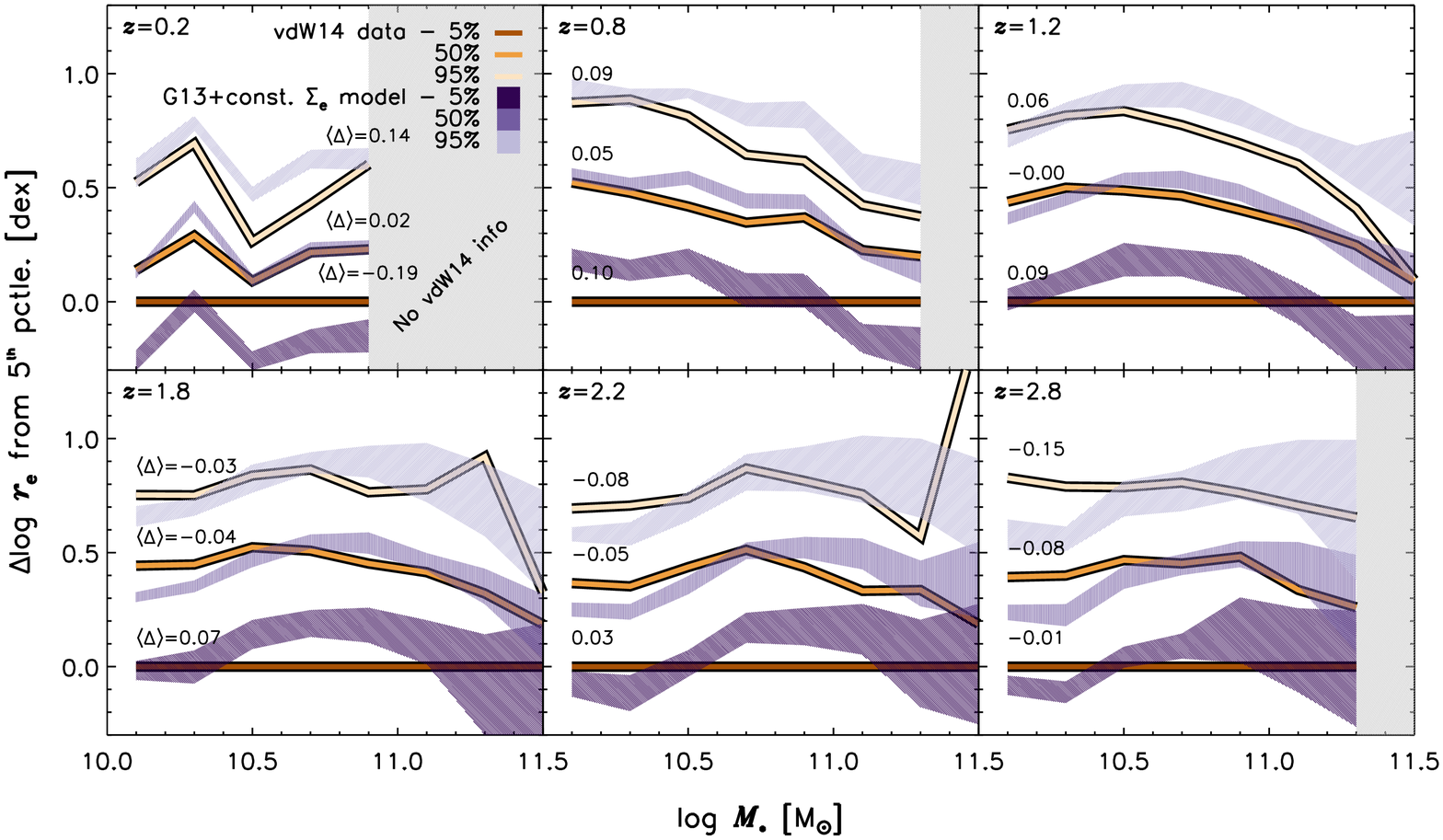}
\vspace{-0.5em}
\caption{{\bfb {\it Top}:} The $z\lesssim3$ size--mass relation implied by constant-$\dens$ growth and G13 lognormal SFHs compared to vdW14 data. {\bfb {\it Bottom}: Model (purple) vs.\ data (orange) size quantiles at fixed $(\Mstel,z)$ relative to the data's 5$^{\rm th}$ percentile (i.e., smallest galaxies; bands show 2\,$\sigma$ model credibility range).} At $z=0$ ({\it top left}), blue/red points show starforming/passive model galaxies ($\log\ssfr\gtrless-11$) with dashes showing linear fits. The model is normalized only to the midpoint of the $z=0$ starforming relation (solid blue line; orange line shows local passive relation; see text). Grey points show model projections at earlier epochs. Above $\log M_{*,0}=10$ and the HST resolution limit, these span/coincide {\bfb well} with vdW14 {\it UVJ}-coded ``late-''/``early-type'' data and fits (blue/orange points/lines showing 1\,$\sigma$ scatter), {\bfb with quantiles never offset by $>$0.2 dex and typically less than 0.1}. Hence, no compaction is necessary to reach the compact red galaxy regime---$\log(\Mstel, r_{e})\approx(10.7, 0)$ at $z\gtrsim2$---{\bfb though the bottom panels reveal the model to underproduce these systems (see Appendix \ref{sec:AA}; Figure \ref{fig:g13CANDELS})}. Notably, {\bfb this level of} agreement over $\Delta t\sim$11 Gyr is achieved {\bfb with all model galaxies moving {\it across} the red/blue loci, and {\it zero} moving along them} (solid arrow at {\it top middle}; dashed arrows show merger effects, which are neglected). The size--mass relation's behavior can thus arise from larger galaxies forming later at lower global densities, and thus naturally being bluer than smaller, earlier-forming systems (see $z_{0}$-shaded circles at {\it top}; dark = old, light = young).}%
\label{fig:sizeMass}
\end{figure*}

We set $\eta$ once at $z=0$, thus (deliberately) ignoring any time dependence in the above phenomena. We do so by matching the modeled and measured starforming {\bfb galaxy sizes at} $\log\Mstel=10.5$ (Figure \ref{fig:sizeMass}, {\it top left}), calibrating to the mean of vdW14's $z=0.25$ ``late-type,''\footnote{\url{http://www.mpia.de/homes/vdwel/3dhstcandels.html} (all fields), with masses/colors from \url{http://3dhst.research.yale.edu/Data.php} \citep{Brammer12, Skelton14, Momcheva16}.} and \citet[][]{Mosleh13}'s $z=0$ $\log\ssfr>-11$ relations (see their Tables 1). The latter is decurcularized using vdW14's $\langle\log(a_{e}/r_{e})\rangle = 0.15$. More sophisticated analyses could probe ellipticity effects, but we assume $r_{e}=a_{e}$ below.

We adopt $\log M_{*,0}=10$; model sizes are undefined below this mass. While free, this choice is not quite arbitrary. It is (1) G13's $z=0$ limit (G13); (2) near where the $z=0$ SFR--$\Mstel$ relation's slope breaks {\bfb below unity} \citep[][]{Salim07,Whitaker14}; (3) the mass above which galaxies have bulges/dense central {\bfb structures} \citep[e.g.,][]{Lang14,Abramson14a}; (4) the mass below which galaxy sizes are indeed largely (though not entirely) mass-independent (vdW14; M17). There is room to argue about $M_{*,0}$ (Section \ref{sec:discussion}), but given our aim and the model's deliberate over-simplicity, these facts suggest $\log M_{*,0}=10$ is reasonable.

This model differs from LC16's in several key ways. Foremost, it applies knowledge of global conditions at only one epoch---$z_{0}$---linking size growth ``ballistically'' to mass at all other times via a different scaling---$r_{e}\propto\Mstel^{1/2}$ vs.\ $\Mstel^{1/3}/(1+z)$. It can do this because it relies on SFHs with {\bfb unique geometries}, not a {\bfb universal} mean $\ssfr(\Mstel;\, t)$ law. Consequently, the model requires no explicit quenching: SFHs simply fall absent any mechanistic prescription (LC16 use mass- {\bfb and environment-}dependent quenching probabilities; a penalty is that we cannot separate centrals and satellites). Also, we make no explicit assumptions about mass profiles (LC16 assume exponentials). {\it We do not claim that our model is ``better,''} only that it represents a meaningful, independent test whose results bear at least on the data's discriminating power.

{\bfb Finally, though addressed in Section \ref{sec:discussion}, we are explicitly not interested in {\it post-}quenching density evolution; i.e., the size growth of individual passive galaxies over time. We thus neglect mergers, which probably drive this phase \citep[e.g.,][]{Newman12a,Morishita16,Belli17}. We comment on the effects of this choice where necessary.}

\subsection{\bfb Results}
\label{sec:modelResults}

Figure \ref{fig:sizeMass}'s {\bfb top panels} show the G13 + constant-$\dens$ predictions for the $(\Mstel, r_{e})$ plane at $z\lesssim3$ {\bfb overlaid on vdW14's data}. {\bfb Bottom panels show size quantiles at fixed mass and time relative to the smallest observed sources (5$^{\rm th}$ pctle.).} Qualitatively and quantitatively, the model reproduces the data well. 

At $z=0$, when split at $\log\ssfr = -11$ into starforming and passive galaxies, blue model galaxies are larger at fixed-mass and exhibit a shallower mass-dependence than their red counterparts ($d\log r_{e}/d\log\Mstel=0.22\pm0.02$ vs.\ $0.40\pm0.02$; results are robust to reasonable choices). {\bfb Meanwhile, red model galaxies} avoid small sizes at high-masses in {\bfb fair} agreement with data descriptions. Indeed, the model blue locus' slope precisely matches the data's {\bfb (vdW14 find $0.25\pm0.02$), as does its median and upper-limb at $z\gtrsim0.8$}. While the model $z=0$ red locus' slope is slightly too shallow, {\bfb including dry merging would} steepen this trend by moving galaxies along lines of $d\log r_{e}/d\log\Mstel = 1$--2 (\citealt{Bezanson09,Hopkins09b,Naab09}).

{\bfb This finding addresses the first question posed in Section \ref{sec:modelResults}: The size--mass relation can be shallower than the constant-$\dens$ slope of 0.5 along which individual galaxies evolve because it is a convolution of tracks launching with larger mean $y$-intercepts (lower-$\dens$) at later times. When projected at any epoch, this stretches the locus horizontally relative to any of its constituents' paths. Hence the discrepancy between galaxy trajectories and loci slopes is superficial.}

The large dots in Figure \ref{fig:sizeMass}, {\it top}---colored by $z_{0}$ for a random subset of SFHs---illustrate the age--size covariance. As expected, and bearing a strong resemblance to spectroscopic and photometric age estimates (\citealt[][]{Valentinuzzi10a}; M17), small systems are redder because they are older, hence further from their peak star formation. Conversely, larger galaxies are bluer because they are younger \citep{Carollo13}. All trends might reflect only the fact that earlier-forming galaxies had their SFHs sped by globally higher densities (Figure \ref{fig:densEvo}; \citealt{Fagioli16, Williams17}). 

The model's {\bfb general} fidelity extends to all other redshifts where vdW14 data exist to test it. {\bfb As Figure \ref{fig:sizeMass}, {\it bottom}, shows, basic G13 + constant-$\dens$ growth predicts $r_{e}(\Mstel)$ quantiles to within $0.2$\,dex, with mean offsets typically less than 0.1. As neither G13 nor constant-$\dens$ growth was conceptualized to describe these data at all quantitatively, we find this agreement striking. It also addresses Section \ref{sec:modelResults}'s second question.

As revealed by the $z>1.2$ panels,} {\it constant-$\dens$ growth can populate the compact galaxy regime}---$\log(\Mstel, r_{e})=(>10.5,0)$. {\bfb Indeed, it does so with some passive systems when defined by a 0.6\,dex or 2\,$\sigma$ cut below $\langle\ssfr(t)\rangle$ (not shown; see below). These systems have $r_{e,0}\sim0.3$\,kpc, near recent high-$z$ estimates \citep[][who find $r_{e}\propto L^{0.5}$ for $z\gtrsim6$ sources, similar to our scenario]{Bouwens17}.} Hence, no compaction is needed {\it to create} such small, passive, high-$z$ objects \citep[concurring with][]{vanDokkum15}. {\bfb That said, the bottom row of Figure \ref{fig:sizeMass}, {\it bottom}, shows that the model does underproduce them: predicted 5$^{\rm th}$ pctle.~sizes are $\sim$0.2\,dex larger than the data suggest. Indeed, $\log\Mstel\sim10.7$ model passive systems are largely responsible for this offset, being up to 2$\times$ too large at $z<2.5$ (Figure \ref{fig:meanSizeZ}, {\it left}). 

While compaction could resolve this tension, so could two compaction-free alternatives.

Principally, we have not correlated the scatter in $r_{e,0}$ (i.e., angular momentum) with SFH properties (half-mass time, width). {\bfb Beyond modulating the number of super high-$\dens$ galaxies, this over mixes the population and ensures we will get $r_{e}(\Mstel,\ssfr;\,t)$ wrong. (This is why we do not plot the model split by $\ssfr$ at $z>0$; it would illustrate $r_{e,0}$ assignment, not constant-$\dens$ growth.)} In reality, scatter in $r_{e,0}$ almost certainly {\it is} correlated with SFH properties \citep[e.g.,][]{Cortese16}, and so $\ssfr(t)$. Studying such correlations is beyond our aims, but---given assembly bias---it is likely such that the fastest-forming, earliest-quenching galaxies arose in overdensities, and so were smaller than equal-mass field galaxies (\citealt{Poggianti13}; M17). This would ``purify'' the compact galaxy population {\bfb into more mature (redder) objects}. {\bfb Appendix \ref{sec:AC} simulates this correlation, which removes any tension in the sizes of red galaxies at $z<3$ (Figure \ref{fig:meanSizeZ}, {\it right}).}

Another satisfactory, compaction-free solution is to allow some model systems to grow at constant $\rho$ \citep[$d\log r_{e}\propto1/3\,d\log\Mstel$;][]{vanDokkum15}, thereby remaining smaller at any $\Mstel$. Since the densest systems comprise $\lesssim$5\% of all galaxies and our basic model accounts for a third to half of them (Figure \ref{fig:g13CANDELS}), adding even a small number of these tracks might resolve all tension. While quantitatively different from our null hypothesis, it is obviously similar in spirit, and seems reasonable given the more spheroidal nature of red galaxies. A}gain, {\bfb our central claim is not} that all galaxies truly evolve at fixed $\dens$, but that the data do not imply compaction.

Section \ref{sec:discussion} presents other remedial modifications. {\bfb Regardless, G13+constant $\dens$ growth {\it as modeled} accounts for the evolution of $\gtrsim$90\% of objects at at least $z\lesssim2$ (Appendix \ref{sec:AB}).}

{\bfb While there is ample room for further exploration, and we encourage others to test any of the above statements, we take the test above as numerical support for the plausibility of} Section \ref{sec:dataResults}'s main points: Absent mergers, compaction, or even explicit quenching, {\bfb and with no tuning beyond $M_{*,0}$ and {\it random} scatter in $r_{e,0}(M_{*,0})$}, once the size--mass relation is reinterpreted as a size--{\it time} relation, the need for rapid pre-quenching density increases largely (if not entirely) disappears. We could thus posit that fixed-$\dens$ growth actually describes most galaxies (at least before quenching), or that the data {\bfb studied} here---numbers, sizes, and colors of galaxies as functions of $\dens$, mass, and time over the past 11\,Gyr---are not very constraining. {\bfb Either way,} though it may happen, these data do not {\it imply} that red galaxies descend from blue galaxies leaping away from their peers towards anomalously high densities. Support for that scenario must be found elsewhere. 


\section{Discussion}
\label{sec:discussion}

Amplifying \citet{Poggianti13,Poggianti13b}---whose spectroscopy demonstrates the slow evolution of the high-$\dens$ population---and LC16---whose $r_{e}\propto\Mstel^{1/3}/(1+z)$ model also suggests $\dens$ trends reflect fundamentally mass- and time-dependent physics---our results highlight the challenge of inferring quenching mechanisms from correlations between galaxy density and quiescence, especially as a null hypothesis of constant-$\dens$ aging also fits the facts. 

Of course, absence of evidence is not evidence of absence. {\it We do not contend that galaxies truly evolve only at fixed $\dens$}: just that this predictive null scenario ought to be falsified and its failures identified before invoking compaction quenching. The caveats of our analysis may provide ways forward.

First, the small areal coverage of the data ($\sim$70 arcmin$^{2}$) may be a concern. Yet, repeating the analysis in Section \ref{sec:advancedResults} using the full suite of vdW14 CANDELS data---$\sim$900 arcmin$^{2}$ but complete to only $\log\Mstel\geq10$---changes none of our conclusions: As shown in Appendix \ref{sec:AA}, Figures \ref{fig:CANDELS} and \ref{fig:sig1CheckCANDELS}, growth of the total $\dens$ function is never more rapid at high- relative to low-$\dens$, and the number of lower-$z$ red galaxies at fixed-$\dens$ never exceeds that of all older galaxies, accounting for quenching times. Hence, there is no {\bfb empirical} evidence of high-$\dens$ (red) systems being drawn from a low-$\dens$ reservoir.

Another concern could be that accurately assessing rest-frame $V-J$ colors at $z\gtrsim1$ may require longer-wavelength photometry than we have used. This could affect the inferred abundance of quiescent galaxies at those epochs. Yet, CANDELS used such data \citep{Skelton14} and, as just discussed, results derived therefrom agree with our assessments. Further, tests show that M17's SED fitting (upon which we rely) tends to {\it overestimate} {\it UVJ}-quiescent galaxy counts by perhaps 50\% relative to assessments using supplementary $K_{\rm s}$ or 3.6$\,\micron$ coverage \citep[][]{Castellano16}, and only at $z<1$. Thus, if anything, our conclusions are conservative. 

In terms of physical counterarguments, principally, minor mergers, adiabatic expansion, and stellar mass loss could dilute the number of high-density red galaxies by pushing some compaction-quenched systems back to lower $\dens$ \citep{Bezanson09,Newman12a,Poggianti13b}. Simulations suggest these phenomena can lower $\dens$ by $\sim$10$\times$ \citep{Naab09,Ceverino15}. {\bfb Relatedly}, red galaxies are reported to grow by factors of $\sim$3--5 in $r_{e}$ at fixed mass over the interval probed \citep[][but see below]{Trujillo07}, which would also lead to large declines in $\dens$.

Yet, for these effects to hide compaction quenching and preserve the signal in Figure \ref{fig:histograms}, {\it right}, the timescales must match: The flow of red galaxies out of high-$\dens$ bins must balance that of blue galaxies into them. Indeed, since there are many more blue galaxies to move right than red ones to move left, the rates of $\dens$-reducing phenomena should be faster than (or tuned to) the quenching timescale. So, progenitors cannot just be dense, but must grow rapidly once quenched, perhaps tripling in size---but not mass---in the 2.5 Gyr between $z\sim2$ and $z\sim1$ \citep[][]{Yano16}.

The general size growth of red galaxies is measured across the entire redshift interval probed, so, in the mean, this process is too slow. Moreover, our null scenario of density-accelerated aging and the LC16 $r_{e}\propto\Mstel^{1/3}/(1+z)$ model naturally explain this as {\bfb ``progenitor bias''}---the addition of lower-density red galaxies descending at later times from lower-density blue progenitors quenched by $\dens$-independent phenomena. Supporting this, \citet{Poggianti13} find red galaxies at fixed mass to grow by $\lesssim2\times$ once stellar age is accounted for. Given the 0.5\,dex $\dens$ bins used here, cohorts of simultaneously quenched red galaxies might thus shift one bin left, leaving our conclusions intact unless migration was a strong function of $\dens$.

On an individual galaxy basis, since quenching times are short{\bfb er} at high-$z$ \citep[$\tau_{\rm Q}=0.35\, t<1.5$ Gyr at $z\geq1.5$, e.g.;][]{Zolotov15}, mergers and adiabatic expansion seem unlikely to drive large density reductions: there are too few of the former and they both take too long \citep[][Figure 3]{Newman12a,Nipoti12, Sonnenfeld14,Barro15}. To compensate, $\tau_{\rm Q}$ could be raised, but at $\gtrsim2$ Gyr it becomes difficult to disentangle $\dens$-quenching from strangulation or gas exhaustion \citep[e.g.,][]{Larson80,Peng15}. If not the aging ``agents'' in the null, fixed-$\dens$ hypothesis, these processes may be more sensitive, e.g., to environment than $\dens$ \citep{Wetzel13}. Indeed, $z\lesssim0.6$ clusters host about triple the number of dense, $\log\Mstel\gtrsim9$ galaxies as the field (\citealt{Poggianti13}; M17), so environmental effects are likely non-negligible for many high-$z$ dense objects.

Now, in fairness, in its current form, the G13 + constant $\dens$ model will not produce the {\bfb absolute smallest} $z>1$ red galaxies $\log(\Mstel, r_{e}) = (10.5,<0)$, assuming the under-resolved measurements in Figure \ref{fig:sizeMass}, {\it top}, are accurate. This is likely due to model SFHs reaching $\log M_{*,0}=10$ at the wrong time to later alight in the correct part of $(\Mstel,r_{e}; z)$ space given their SFRs. {\bfb Beyond the moves discussed in Section \ref{sec:modelResults},} making $M_{*,0}$ a function of time might alleviate this issue. Simply lowering $M_{*,0}$ will not work: While it leads to smaller galaxies, it also produces a size--mass relation with a slope close to 0.5: though clocks start earlier, such that $r_{e,0}(\Mstel)$ drops due to globally higher densities, the SFHs are also further from their terminal masses. Hence, $dr/d\Mstel$ is integrated over a larger domain, stretching the constant-$\dens$ tracks into $\sim$parallel lines.

To maintain the correct $r_{e}(\Mstel)$ shape while generating $\lesssim$kpc massive galaxies at $z\gtrsim1$, $M_{*,0}$ must be bent toward {\it higher} masses at earlier times, causing objects to start nearer their destination at epochs when their SFHs are also closer to completion. While beyond the scope of this work, such a treatment is consistent with the evolution of the break in the $\sfr$--$\Mstel$ relation \citep{Whitaker12, Schreiber15} and so might be meaningful; {\bfb we encourage others to see if the above can account for the small number/overabundance of hyper-dense systems our first attempt did not produce.}

In sum, rapid secular expansion seems the best out for compaction/density-triggered quenching. Since $\tau_{\rm Q}\gg t_{\rm dyn}$ at $z\lesssim3$, timescales would accommodate it, and simulations suggest it is possible if star formation ends in a large burst \citep{elBadry16}. The mechanism seems most active at $\log\Mstel\leq9.6$---about a dex below most high-$z$ red galaxies---but there is some evidence at $z\gtrsim1.5$ and $\log\Mstel\gtrsim10.5$ that the largest red systems are also the {\it oldest} \citep{Yano16} {\bfb or the youngest red systems are the smallest \citep[][]{Almaini17}}. If confirmed using larger {\bfb spectroscopic} samples and shown not to reflect, e.g., merger-driven rejuvenations of previously red galaxies ({\bfb perhaps} if the poststarbursts are not too dusty), this would be a ``smoking gun'' of compaction-triggered quenching, ruling out {\it exclusively} constant-$\dens$ evolution (but see LC16). If simulations support rapid expansion, non-structural predictions such as ages, metallicities, or $\alpha$-abundances would aid observers in testing such hypotheses.


\section{Summary}
\label{sec:summary}

\begin{figure*}[t!]
\centering
\includegraphics[width = 0.45\linewidth, trim = 0.5cm 0cm -0.5cm 0cm]{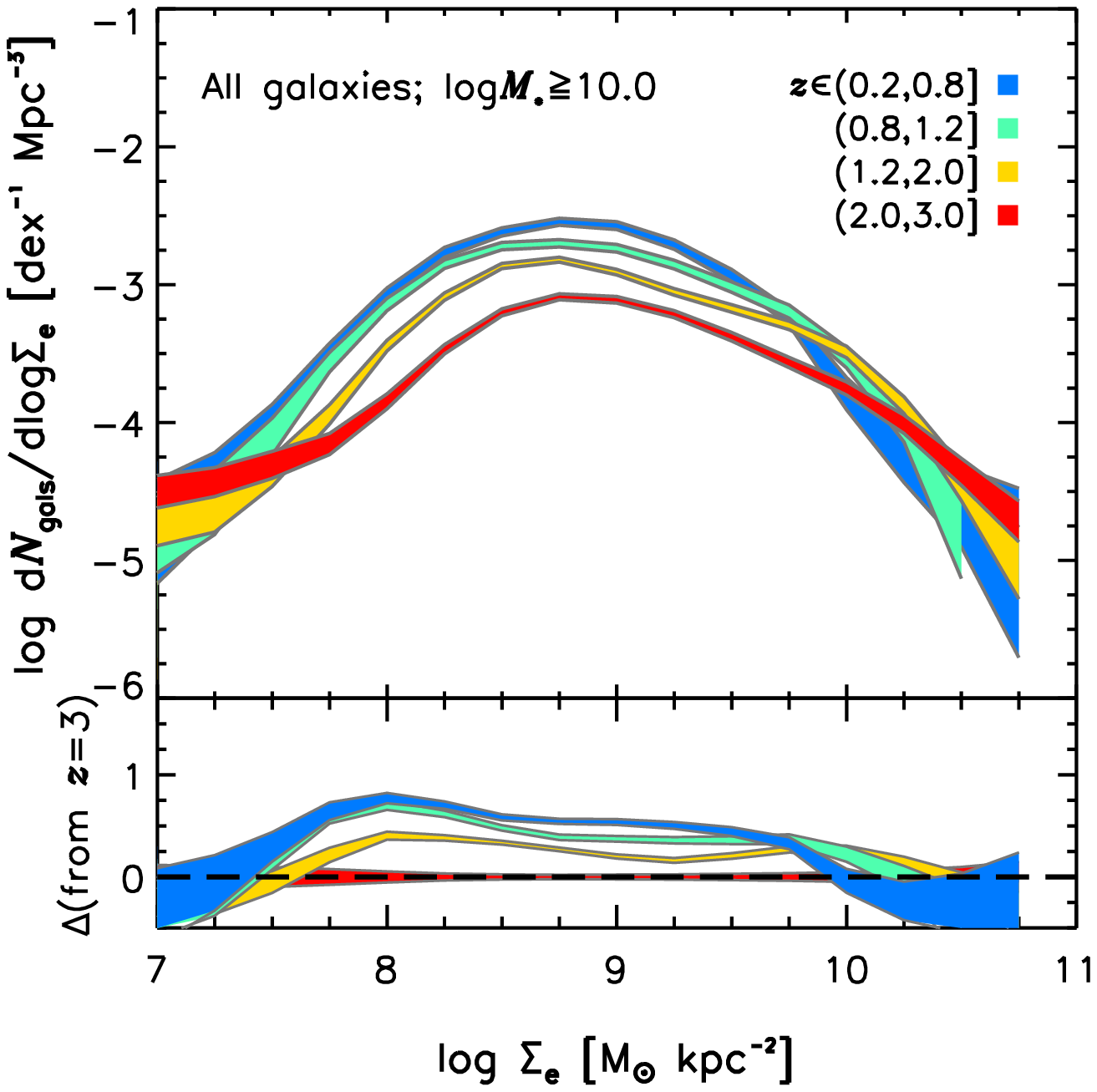}
\includegraphics[width = 0.45\linewidth, trim = -0.5cm -0cm 0.5cm 0cm]{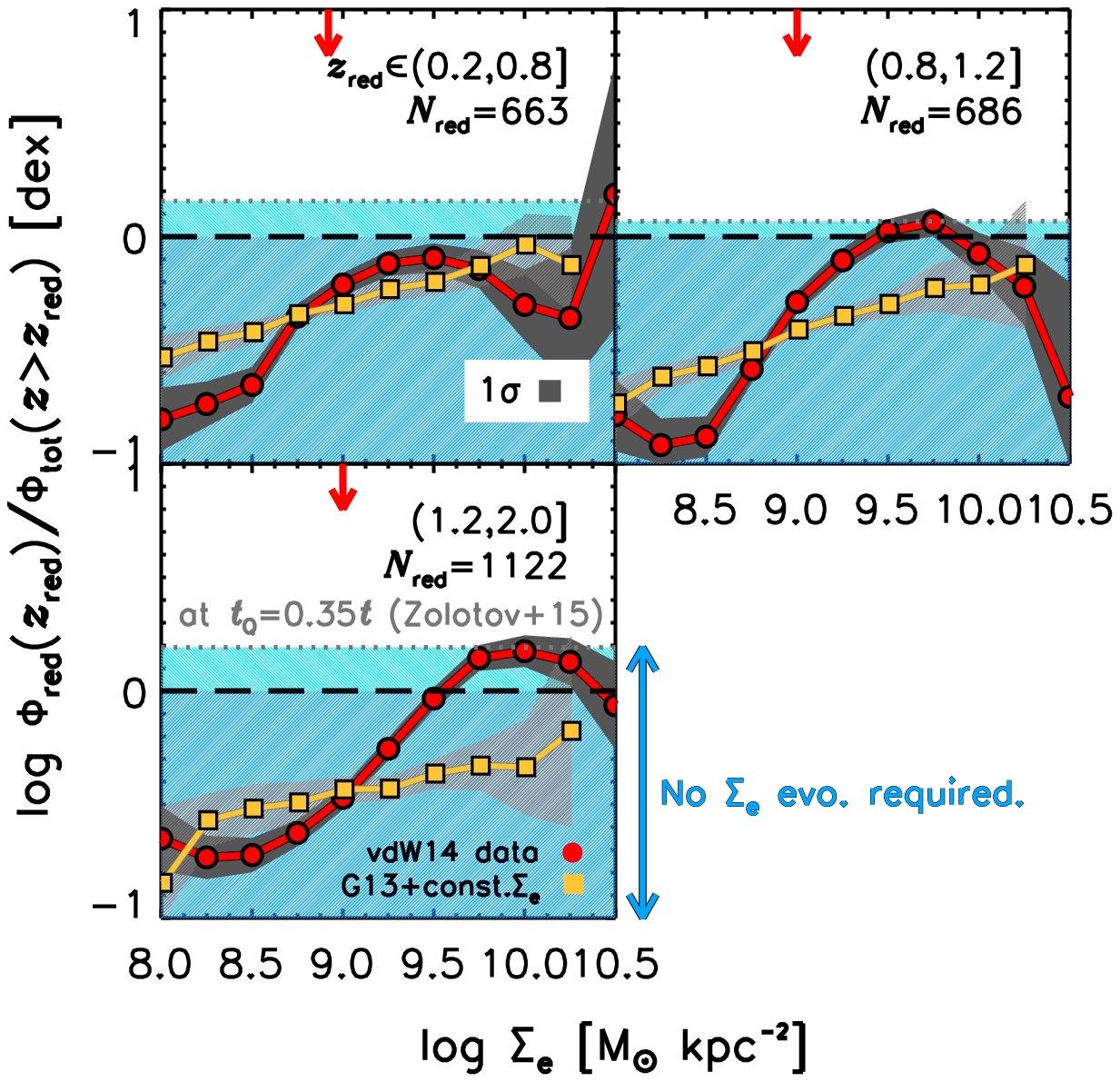}
\caption{Figure \ref{fig:histograms} reproduced using wider-but-shallower CANDELS data. Results are consistent with those based on deeper XDF data. {\bfb Yellow lines at {\it right} show G13 + constant $\dens$ model predictions assuming passive galaxies have $\log\ssfr(t)/\langle\ssfr(t)\rangle<-0.6$\,dex, or $\sim$2$\sigma$ below $\langle\sfr(\Mstel;\,t)\rangle$.}}
\label{fig:CANDELS}
\end{figure*}

Using the deepest HST data obtained, we show that galaxy colors, stellar mass surface densities ($\dens\equiv\Mstel/2\pi r_{e}^{2}$), sizes, and abundances at $z\lesssim3$ and $\logM\geq\compl$ are consistent with a scenario in which all systems evolve from blue to red at roughly fixed $\dens$ at rates correlated with that quantity. Though it may occur, there is no requirement that blue galaxies quench by compaction, evolving dramatically in $\dens$ (beyond some critical threshold). Specifically:
\benum
	\item There is no preferred density at which blue galaxies turn red (Figure \ref{fig:densEvo}). Rather, this process occurs at all $6.5\leq\log\dens/\Msun\,\kpc^{-2}\leq10$, with denser systems reddening earlier/faster than less-dense ones.
	\item The number of red galaxies never exceeds that of all equal-$\dens$ galaxies at earlier times (Figures \ref{fig:histograms}, \ref{fig:CANDELS}, {\it right}), which also have consistent masses and central-kiloparsec densities (Figures \ref{fig:sig1Check}, \ref{fig:sig1CheckCANDELS}). There is no suggestion of a large influx of once-lower-density blue galaxies into the high-density red population.
	\item A simple model based on G13 lognormal star formation histories where galaxies never increase in $\dens$ reproduces the $z=0$ size--mass relation of blue and red galaxies, and the evolution of the entire locus at $\log\Mstel\gtrsim10$ and $z\lesssim3$ (Figures \ref{fig:sizeMass}, \ref{fig:g13CANDELS},  \ref{fig:meanSizeZ}).
\eenum

While we can neither {\bfb prove that compaction quenching {\it never} occurs, nor} rule out all scenarios that might mask {\bfb it} (e.g., rapid secular expansion), the most straightforward (minor mergers) seem unlikely given timescale requirements. Thus, a null scenario in which, {\bfb pre-quenching}, galaxies evolve at roughly constant $\dens$---with denser galaxies aging more rapidly from blue to red via gas exhaustion or other Hubble-timescale processes, which are by definition rapid at high-$z$---seems equally plausible. Future investigators should falsify this and predict non-structural characteristics of galaxies undergoing compaction quenching (e.g., ages, $\alpha$-enhancements) to aid observers in testing the implied mechanism(s).

\section*{}
\noindent We thank T.\ Treu, B.\ Poggianti, B.\ Vulcani, O.\ Almaini, D.\ Masters, M.\ Kriek, A.\ Wetzel, {\it Galpath2016} attendees, and our anonymous referee for helpful insights. TM acknowledges a Japanese Ministry of Education, Culture, Sports, Science and Technology Grant-in-Aid for Scientific Research (26-3871), and a Japan Society for the Promotion of Science research fellowship for young scientists. GLASS (HST GO-13459) is supported by NASA through a grant from STScI operated by AURA under contract NAS 5-26555.


\appendix
\section{A: Sample Size}
\label{sec:AA}

\begin{figure}[t!]
\centering
\includegraphics[width = 0.95\linewidth, trim = 0.5cm 0cm -0.5cm 0cm]{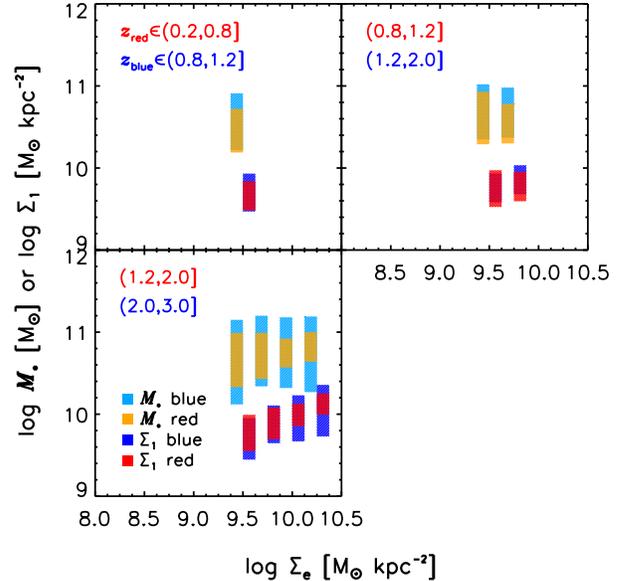}
\caption{A reproduction of Figure \ref{fig:sig1Check} based on wider-but-shallower CANDELS data. Results are fully consistent with those from deeper XDF data.}
\label{fig:sig1CheckCANDELS}
\end{figure}

Figures \ref{fig:CANDELS}, \ref{fig:sig1CheckCANDELS} reproduce Figures \ref{fig:histograms}, \ref{fig:sig1Check}, \resp, based on wider and shallower data from CANDELS (vdW14). As discussed in Section \ref{sec:advancedResults}, Figure \ref{fig:CANDELS}, {\it left}, shows that, while galaxy abundances have increased at most $\dens$ since $z\sim3$, the growth at high-$\dens$ is never faster than that at low-$\dens$. If anything, the trend goes in the opposite sense. Thus, there is no evidence {\bfb in this larger sample} for an anomalous number of high-$\dens$ galaxies drawn from a reservoir of previously low-$\dens$ galaxies via rapid shrinking/compaction events. (Negative growth at the highest-$\dens$ probably reflect post-quenching dry mergers, which are too slow to mask compaction; see Section \ref{sec:discussion}.)

Figures \ref{fig:CANDELS}, {\it right}, and \ref{fig:sig1CheckCANDELS} show that all statements hold once these data are split into starforming/quiescent galaxies using {\it UVJ} criteria: At no redshift does the $\dens$ function of quiescent galaxies (red lines) exceed that of all older equal-$\dens$ galaxies, accounting for $\tau_{\rm Q}$ (dotted horizontal lines), {\bfb and $\Mstel$ and $\Sigma_{1}$ are consistent in $\dens$ bins with substantial inter-interval quenching}. Indeed, quiescent galaxies never exceed all galaxies in the previous redshift bin---typically separated by {\it more} than a quenching time---except at $\log\dens\sim10$ at $z\sim1.2$--2 (descending from galaxies at $z\sim2$--3; note that this $\dens$ is well above the threshold identified by \citet{Whitaker17}). Hence, using a much wider dataset, we still find no evidence for rapidly shrinking blue galaxies to be a meaningful production channel for the quiescent population.

{\bfb Finally, the yellow trends in Figure \ref{fig:CANDELS}, {\it right}, show G13 + const.\ $\dens$ model predictions assuming passive galaxies have $\log(\ssfr/\langle\ssfr\rangle)<-0.6$\,dex ($\sim$2\,$\sigma$ low-side outliers; e.g., \citealt{Speagle14}). This quantitatively captures the data at low- and high-$\dens$, though not the shape of the trend at $z\gtrsim0.8$. To some extent, this reflects a combination of the passive definition and the randomization of initial sizes, which should likely be correlated with SFH features (Section \ref{sec:toyModel}; Appendices \ref{sec:AB}, \ref{sec:AC}). To prove it a reflects failing of the constant-$\dens$ assumption, one must marginalize over all plausible SFH models and boundary conditions. We leave such exploration to future work.} 

\begin{figure*}[t!]
\centering
\includegraphics[width = 0.45\linewidth, trim = 0.5cm 0cm -0.5cm 0cm]{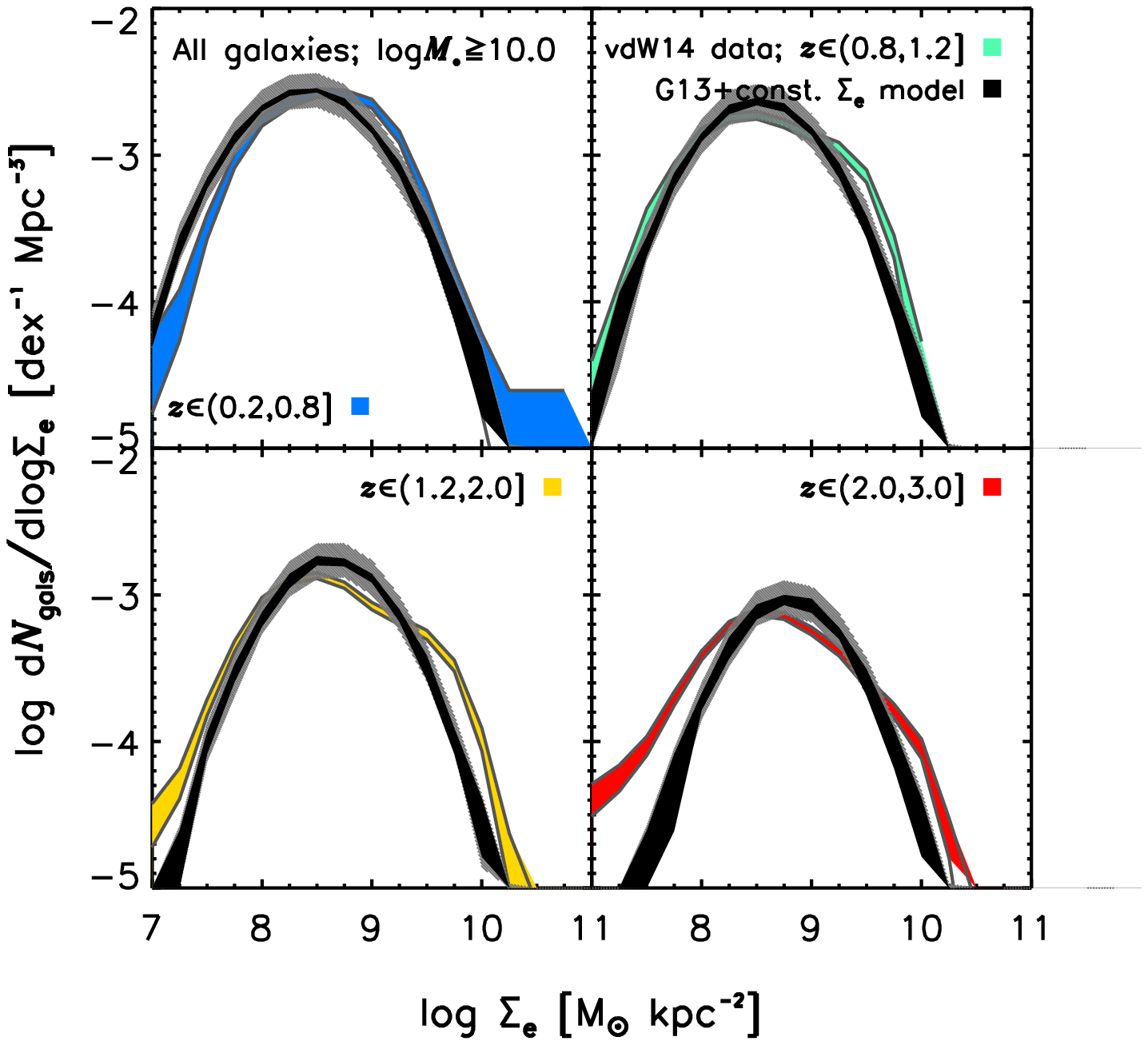}
\includegraphics[width = 0.45\linewidth, trim = 0.5cm 0cm -0.5cm 0cm]{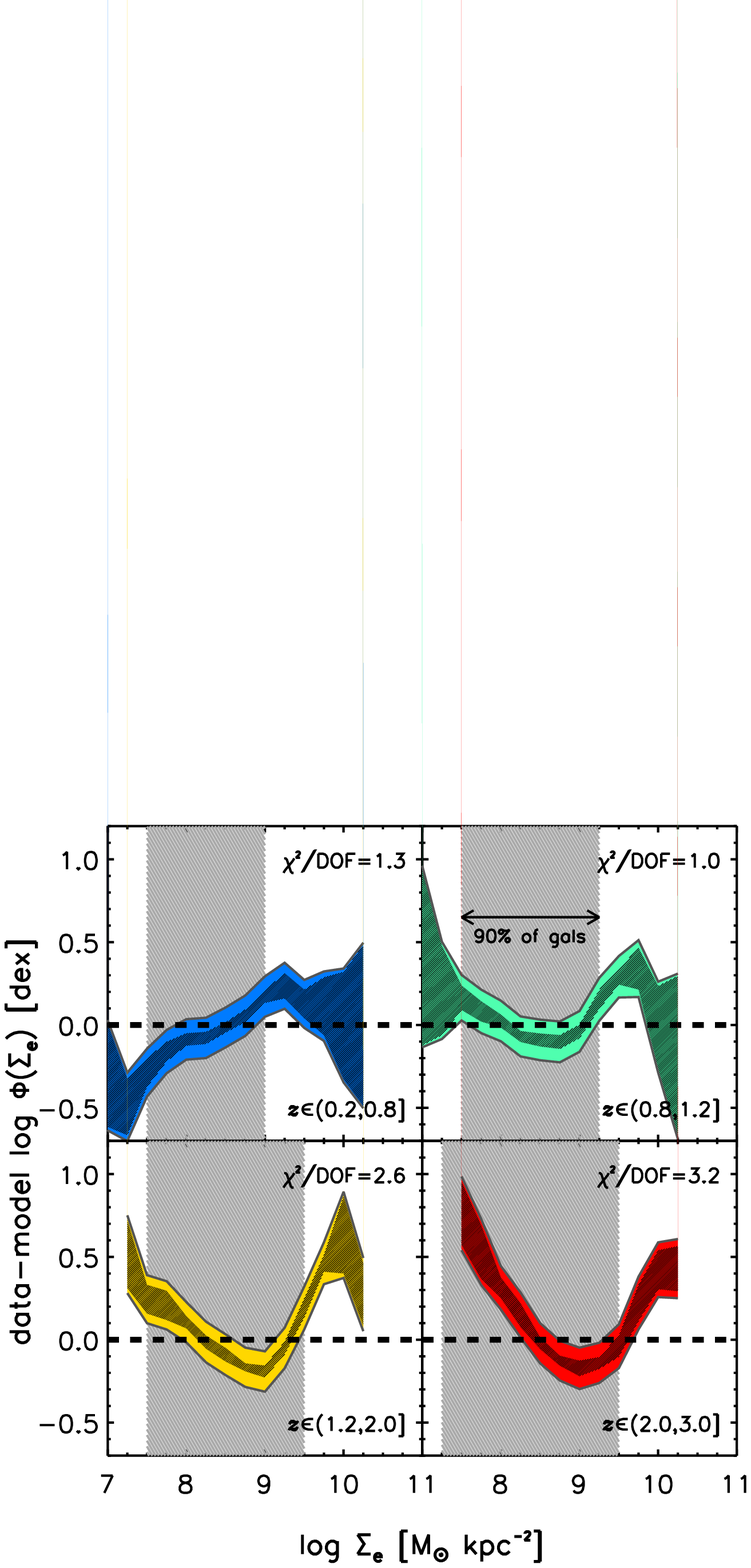}
\caption{\bfb CANDELS data from Figure \ref{fig:CANDELS} split by redshift and replotted using $a_{e}$-based $\dens$ to be consistent with the comparison in Figure \ref{fig:sizeMass}. The $\dens$ functions inferred from that exercise are shown in black shading (1\,$\sigma$ credibility) with normalization uncertainty contributions in grey. The model reaches all densities probed by the data, though tends to underproduce the densest $\sim$5\% of systems by perhaps $\sim$3$\times$ (residuals shown at {\it right}; grey bands denote 5$^{\rm th}$--95$^{\rm th}$ population percentiles). Appendix \ref{sec:AC} suggests retuning the model's initial conditions rectifies this ofset (Section \ref{sec:toyModel}). If not, it may suggest compaction for $\sim$3\% of galaxies.}
\label{fig:g13CANDELS}
\end{figure*}


\section{\bfb B: Further quantitative tests of the Basic G13 + constant-$\dens$ growth model}
\label{sec:AB}

{\bfb Figure \ref{fig:g13CANDELS}, {\it left}, shows Figure \ref{fig:CANDELS}'s vdW14 $\dens$ functions split by redshift and recast using $r_{e}= a_{e}$ for consistency with the comparison in Figure \ref{fig:sizeMass}. G13 + constant $\dens$ model predictions are overplotted as black/grey shaded bands. Band widths reflect 20 re-realizations of Section \ref{sec:modelIngredients}'s procedure with errors in volume normalization---performed only at $z\sim0.5$---added in quadrature. Figure \ref{fig:g13CANDELS}, {\it right}, shows the data$-$model residuals with the $\dens$ interval containing 90\% of galaxies highlighted.

At all $z<2$, the model-predicted abundances are consistent with the data at $\dens$ describing the vast majority of galaxies. Indeed, with a $y$-axis offset determined only at the lowest-$z$, formal $\chi^{2}$ values are reasonable at all epochs. At $z=1.2$--3, the model does significantly underproduce the densest $\sim$5\% of galaxies by a factor of $\sim$3 (true at $\lesssim$2\,$\sigma$ to a factor of $\sim$2 in one or two bins at $z<2$). Of course, it does {\it produce} such galaxies, so it is possible these discrepancies could be mediated by tweaks to the model boundary conditions (Appendix \ref{sec:AC}), which were agnostic to these data (Section \ref{sec:modelResults}). If not---and the purely empirical counterarguments in Section \ref{sec:dataResults} or Appendix \ref{sec:AA} are unpersuasive---these excesses imply an unmodeled source for the densest objects; perhaps compaction. Regardless, the largest discrepancies in Figure \ref{fig:g13CANDELS} are in fact at the lowest-$\dens$ at $z>2$. While perhaps a meaningful model shortcoming, this seems unrelated to compaction, which is invoked to explain the opposite end of the $\dens$ spectrum.}


\begin{figure}[t!]
\centering
\includegraphics[width = 0.95\linewidth, trim = 0cm 0cm 0cm 0cm]{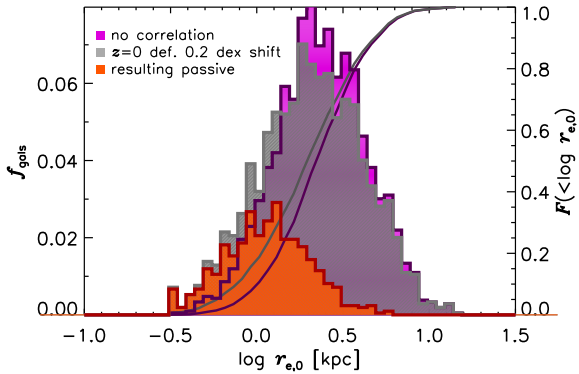}
\caption{\bfb Initial size distributions from the unmodified G13+constant-$\dens$ model (purple), and using modified boundary conditions assuming $z=0$ passive G13 systems have $r_{e,0}$ biased 0.2\,dex (1\,$\sigma$) low (grey). Affected systems are shown in red. This move barely changes the full $r_{e,0}$ distribution, but removes any tension in the predicted vs.\ measured sizes of red galaxies. Solid lines show cumulative distributions for the original and modified models, quantified by the right-hand ordinate.}
\label{fig:shiftRed}
\end{figure}

\begin{figure*}[t!]
\centering
\includegraphics[width = 0.45\linewidth, trim = 0cm 0cm 0cm 0cm]{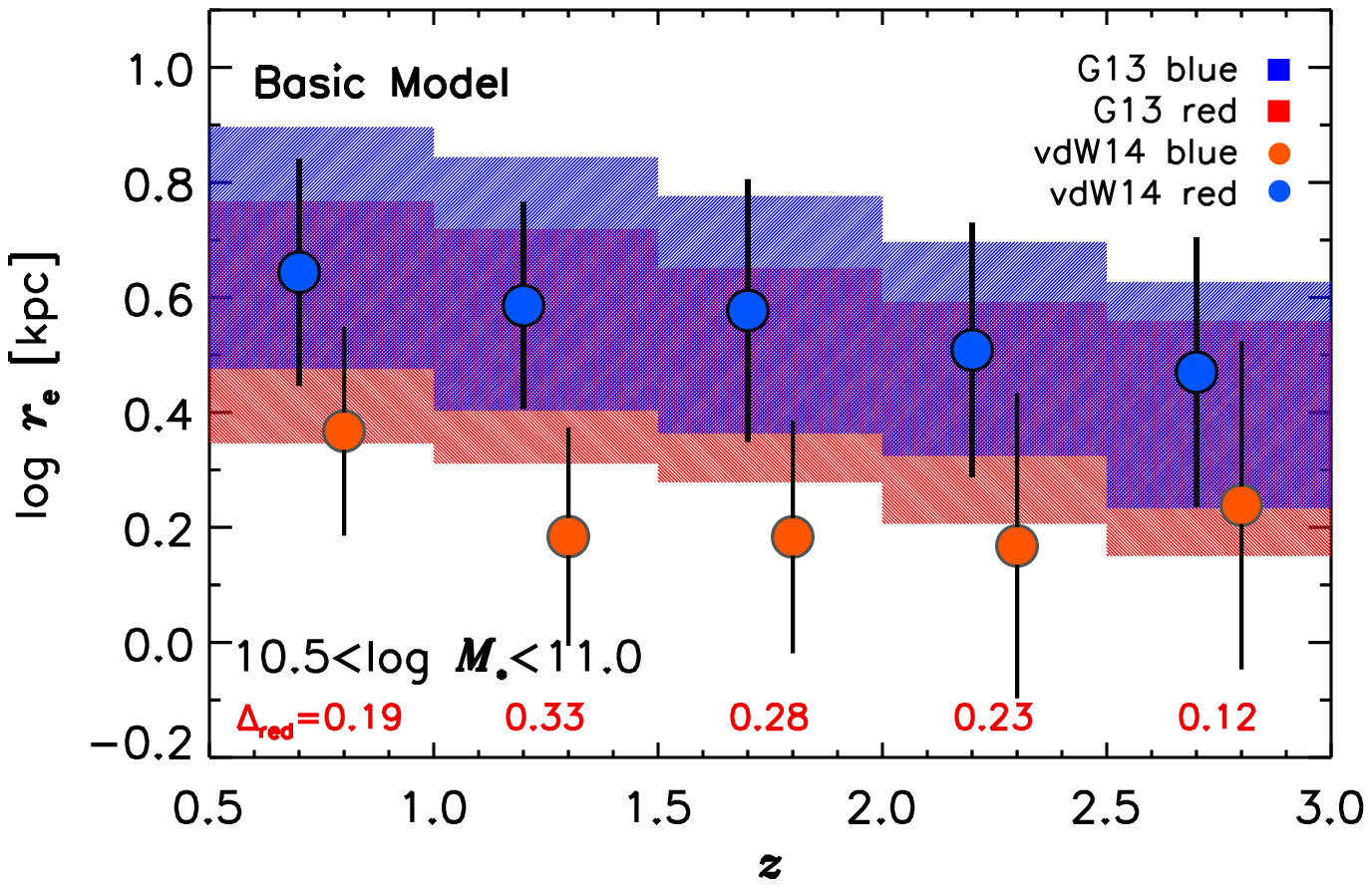}
\includegraphics[width = 0.45\linewidth, trim = 0cm 0cm 0cm 0cm]{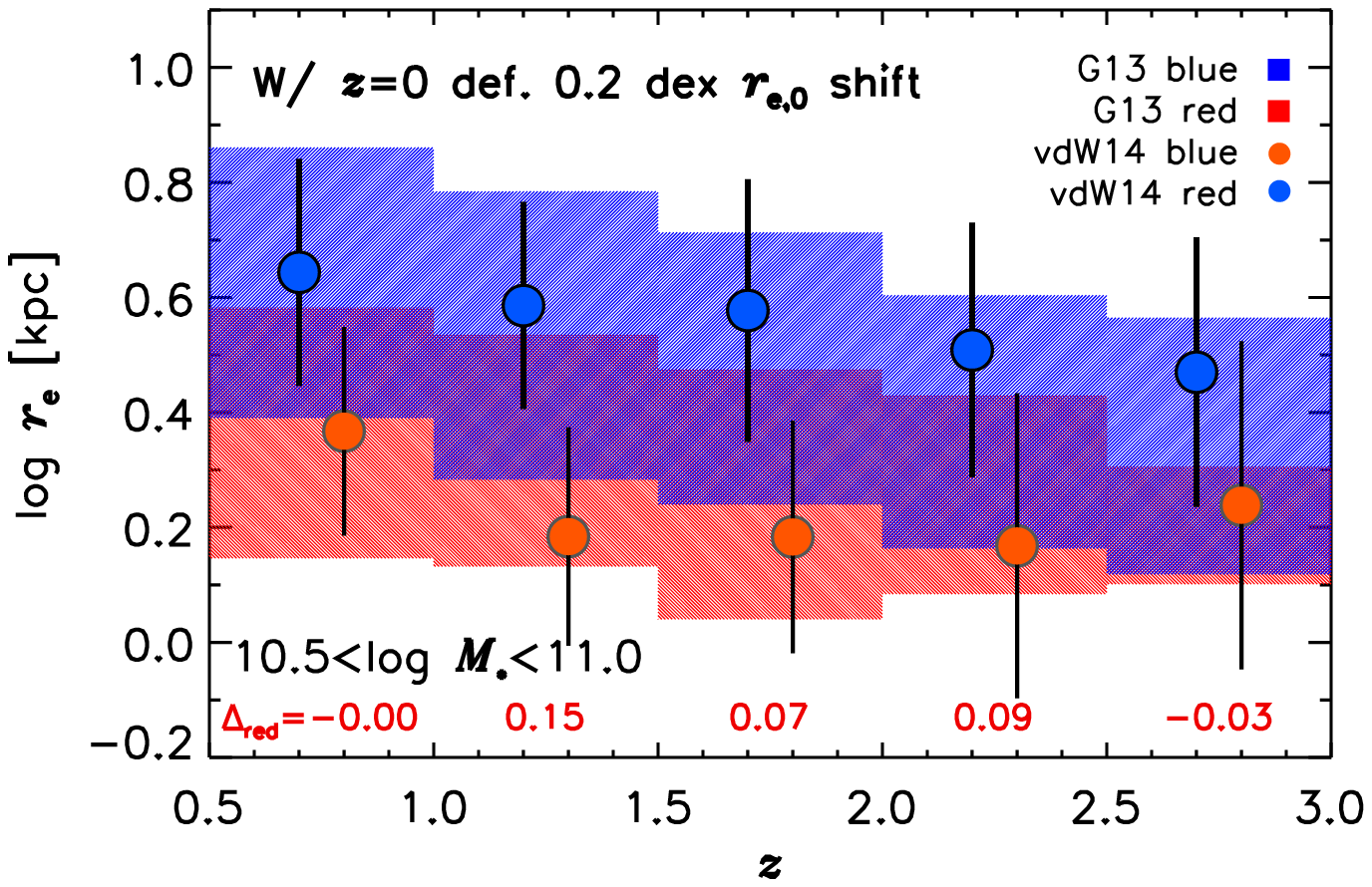}
\caption{\bfb The sizes of $\log\Mstel\sim10.7$ (Milky Way-mass) starforming and passive galaxies at $z<3$. Points show vdW14 {\it UVJ}-defined data, shaded blocks show G13+constant-$\dens$ $\ssfr(t)$-based projections. Error bars/shading denotes 68\% size spreads. Red text along the $x$-axis quantifies the model--data median offsets for red galaxies. The {\it left} panel presents the model in its original form (Section \ref{sec:modelIngredients}), while the {\it right} panel shows results using slightly modified boundary conditions (Figure \ref{fig:shiftRed}). This alleviates almost all tension, suggesting the $r_{e,0}$ assignment is at fault, not the constant-$\dens$/non-compaction size growth framework.}
\label{fig:meanSizeZ}
\end{figure*}

\section{\bfb C: Effects of a simple boundary modification}
\label{sec:AC}

{\bfb As discussed in the text, the boundary conditions of the basic G13+constant-$\dens$ model in Section \ref{sec:modelIngredients} can be modified to ensure that sufficiently small, passive, high-$z$ galaxies are produced without invoking compaction. Principally, instead of being random, the scatter in $r_{e,0}$ can be correlated with SFH parameters, biasing smaller objects towards faster-aging systems. This is is akin to linking red galaxies to dispersion-support/low net-angular momentum, as is observed \citep[e.g.,][]{Fall13,Cortese16}.

A full analysis of such correlations is beyond the scope of this paper, but we provide a reasonable simulation here by simply shifting $r_{e,0}$ by $-$0.2\,dex (1\,$\sigma$) for all G13 galaxies that are passive {\it today}. As Figure \ref{fig:shiftRed} shows, the resulting $r_{e,0}$ distribution is hardly modified from that of the original model. Yet, as Figure \ref{fig:meanSizeZ} shows, all tension in the sizes of, e.g., $\log\Mstel\sim10.7$ G13 passive model galaxies at $z\leq3$---defined by $\ssfr$ {\it at the epoch of observation}---is removed. Such systems are $\lesssim$0.3\,dex larger than vdW14's measurements using the basic model (for well-resolved systems; left panel), but the simply modified version gets them just right (right panel). This is accomplished assuming only constant-$\dens$ growth; we even ensure no galaxies have smaller initial sizes than the minimum produced in the original version. Also, it was not a guaranteed outcome: If all galaxies quenched at $z = 0.1$ in G13, for example, we would still get Figure \ref{fig:meanSizeZ}, {\it right}, wrong. Hence, this simple modification correctly links quenching epochs to sizes. Surely, a more sophisticated treatment would also succeed.}

\bibliographystyle{apj}
\bibliography{/Users/labramson/lit}


\end{document}